\documentclass[12pt]{article}
\usepackage{graphicx}
\usepackage{sidecap}
\usepackage{wrapfig} 
\usepackage{subfig}
\usepackage{float}
\usepackage{amsmath} 
\usepackage{amsfonts} 
\usepackage[mathscr]{eucal}
\begin{document} 
\title{Two-qudit topological phase evolution under dephasing} 
\author{L. E. Oxman$^a$, A. Z. Khoury$^a$, F. C. Lombardo$^b$, and P. I. Villar$^b$ \\ \\
$^a$Instituto de F\'\i sica, Universidade Federal Fluminense, \\ 24210-346 Niter\'oi - RJ, Brasil \\ \\
$^b$Departamento de F\'\i sica {\it Juan Jos\'e
Giambiagi},  FCEyN UBA \\ and IFIBA CONICET-UBA, Facultad de Ciencias Exactas y Naturales, \\
Ciudad Universitaria, Pabell\' on I, 1428 Buenos Aires, Argentina }

\maketitle 
  
\begin{abstract}

In this work, we study a bipartite system composed by a pair of entangled qudits coupled to an environment. 
Initially, we derive a master equation and show how the dynamics can be restricted to a ``diagonal'' sector that includes a maximally entangled state (MES). Next, we solve this equation for mixed qutrit pairs and analyze the $I$-concurrence $C(t)$ for the effective state, which is needed to compute the geometric phase when the initial state is pure. Unlike (locally operated) isolated systems, the coupled system leads to a nontrivial time-dependence, with $C(t)$ generally decaying to zero at asymptotic times. However, when the initial condition gets closer to a MES state, the effective concurrence is more protected against the effects of decoherence, signaling  
a transition to an effective two-qubit MES state at asymptotic times. This transition is also observed in the geometric phase evolution, computed in the kinematic approach.
Finally, we explore the system-environment coupling parameter space  and show the existence of a Weyl symmetry among the various physical quantities. 

\end{abstract}

\section{Introduction}   

It is well known that the history of a cyclic evolution can be retained in the form of a geometric phase (GP). This was first put forward by Pancharatnam in classical optics \cite{Pancharatnam} and later by Berry in Quantum Mechanics \cite{Berry}. Since then, great progress was achieved in this field, ranging from nonadiabatic extensions \cite{Simon-Mukunda} to investigations aimed at implementing geometric quantum computation in NMR (nuclear magnetic resonance) \cite{NMR}, Josephson junctions \cite{JJ}, ion traps \cite{IT}, and quantum dots \cite{QD}. One main goal of quantum computation is to investigate efficient circuits to synthesize quantum operations \cite{nat1,nat2,nat3}. The circuit complexity for qubit systems was investigated within different approaches, including formulations based on the shortest path in a curved geometry \cite{nat5,nat6,nat7,nat8}. Quantum computation schemes were proposed based on either Abelian or non-Abelian GPs. 

Due to its global properties, the geometric phase is propitious to construct fault tolerant quantum gates. In this respect, the GP was shown to be robust against certain external noise sources \cite{refs1,refs2,refs3}.  Interactions play an important role in the realization of some specific operations. As the gates operate slowly compared to the dynamical time scale, they become vulnerable to open system effects and parameter fluctuations that may lead to coherence loss. This motivated the extension of the GP to open quantum systems. 
No matter how weak the coupling to the environment is, the evolution of an open quantum system is eventually plagued by nonunitary features like decoherence and dissipation. Decoherence, in particular, is a quantum effect whereby the system loses its ability to exhibit interference effects.
Following this idea, many authors analyzed how to obtain the GP under the influence of an external environment using different approaches (see \cite{Tong, Tong-err,pra,nos,pau} and references therein).

The understanding of GPs for entangled states is particularly relevant due to potential applications in holonomic quantum computation with spin systems, which provide a plausible design of a solid-state quantum computer. The kinematic approach to the geometric phase for mixed quantal states under nonunitary evolutions was proposed in \cite{Tong,Tong-err}.
The effect of the environment on a bipartite two-level system coupled to an external environment was calculated in Ref. \cite{bipartite}.
The interplay between geometric phases, entanglement and decoherence in a bipartite qubit system was analyzed in Ref. \cite{entanglement}; in particular, the GP correction for an initial maximally entangled state was shown to be null. That is, the phase is built as 
for unitary evolutions, with a stepwise behavior in steps of $\pi$ \cite{MM,M}. This is the case no matter the type or strength of environment to which the system is coupled. In all other cases, not only the GP but also the concurrence is modified by the presence of the environment. 
Topological phases in multiquibit systems were investigated in Ref. \cite{multiqubit}.  

In comparison to the qubit system, $d$-dimensional quantum states (qudits) could be more efficient in quantum applications. With a larger space of states, the qudit algorithms may improve channel capacity \cite{nat9,nat10} and the implementation of quantum gates \cite{nat11,nat12,nat13}, as well as increase security \cite{nat14} and quantum features \cite{nat15}. Qudit systems have also been experimentally realized \cite{nat16,nat17,nat18,nat19,nat20,nat21}.  In Ref. \cite{oxman1}, different sectors of entangled qudits operated by unitary local evolutions were analyzed, identifying geometrical and topological aspects.  The GP was explicitly calculated in terms of the $I$-concurrence introduced in \cite{Iconc},  predicting that for cyclic evolutions in the multiply connected sector of maximally entangled states, it is built in  fractional steps of $2\pi/d$. This sector was identified with group elements in the adjoint representation of $SU(d)$, thus generalizing the $SO(3)$ manifold considered for qubit pairs in \cite{MM,M}. 
This study was later extended to pairs of qudits with general dimensions $d_A$ and $d_B$ \cite{oxman2}. For partially entangled states with different qudit dimensions, the GP can assume continuous values in addition to the fractional phase contribution. In ref. \cite{Asp}, 
the topology of projective subspaces with definite concurrence was further studied and 
GPs were connected with the common elements characterizing the $\mathfrak{su}(d)$ Lie algebra, such as roots and weights. The experimental observation of fractional topological phases with photonic qudits was achieved in Ref. \cite{experiment}.

In the present article, we analyze the effect of a quantum reservoir on a bipartite system formed by a pair of entangled qudits, focusing on the system's geometric phase and related physical quantities under dephasing. The paper is organized as follows. 
In Sec.\ref{cartan}, we present our main algebraic tools. In Sec.\ref{GP},  we briefly review the kinematic approach to compute GPs for a mixed state. In Sec. \ref{master}, we compute the master equation and show how the system dynamics gets restricted to a sector of the reduced density matrix. As an example, we re-obtain the robustness of maximally entangled qubits. In Sec. \ref{master2}, we solve the master equation for two-qutrit states, while in Sec. \ref{concurrence} we discuss the effective concurrence and its relation to the evolution of density matrix eigenvalues. The effect of decoherence on GPs, for an initial qutrit MES state, is discussed in Sec. \ref{GPmes}. In that section, we also explore the system-environment coupling parameter space and show the existence of a Weyl symmetry among the various physical quantities. Finally, in Sec. \ref{conclusiones} we present our conclusions. 
  
\section{Mixed two-qudit states and the Cartan decomposition} 
\label{cartan}   
 
Initially, let us discuss some algebraic tools in Refs. \cite{oxman1, oxman2,Asp} and apply them in the description of mixed states. 
The states of a two-qudit system are naturally expanded as $\psi_{ij}\, |i j\rangle$, $|\ij\rangle = |i\rangle \otimes |j\rangle$, where $|i\rangle$, $i =1,...,d$, is a basis for each individual subsystem. Organizing the expansion coefficients in matrix form $\Psi|_{ij} = \psi_{ij}$, the state can be denoted as $|\Psi \rangle$, and the scalar product is given by,  
\begin{equation}
\langle \Phi | \Psi \rangle = \bar{\phi}_{k l} \psi_{i j}\, \langle k l | i j \rangle = \bar{\phi}_{i j} \psi_{i j} = \Psi|_{i j} {\Phi}^{\dagger} |_{i j} = {\rm Tr}\, (\Psi \Phi^{\dagger})\;.
\end{equation}  
A normalized state is then associated with a matrix $\Psi$, with complex coefficients, such that ${\rm Tr}\, (\Psi \Psi^\dagger)=1$\,. For example, a separable state corresponds to $\Psi_{ij}=a_i b_j$, while a maximally entangled state to $I/\sqrt{d}$, where $I$ is the ($d\times d$) identity matrix. 
In general, we can decompose $\Psi$ into a multiple of the identity plus a traceless piece, which can be expanded using a set of hermitian generators of the Lie algebra of $SU(d)$, $T_A$, $A=1,\dots, d^2-1$\,. 
Here, we will use the normalization ${\rm Tr}\, (T_A T_B) = \delta_{AB}$ and, to simplify the notation, we introduce an index $\bar{A}=0,1, \dots, d^2-1$, defining a basis $T_{\bar{A}}$ for the space of complex matrices, with $T_0 =\frac{I}{\sqrt{d}}$\,,
\begin{equation} 
\Psi =  \psi_{\bar{A}} \, T_{\bar{A}} \makebox[.5in]{,} \psi_{\bar{A}} \in \mathbb{C} \;.
\label{expand} 
\end{equation}
\begin{equation}
|\Psi \rangle = \psi_{\bar{A}}\, |\bar{A} \rangle   \makebox[.5in]{,}  \langle \bar{A}|\bar{B}\rangle =\delta_{\bar{A}\bar{B}} \;,
\label{expand2} 
\end{equation} 
where, as a shorthand notation, we are using $|\bar{A}\rangle = |T_{\bar{A}}\rangle$\,,
\begin{equation}
| T_0 \rangle = \frac{1}{\sqrt{d}}\,\sum_i | i i \rangle\makebox[.5in]{,}| T_A \rangle =  T_A |_{i j} |i j \rangle  \;.
\end{equation}
Equations (\ref{expand}) or (\ref{expand2}) give a natural representation of  two-qudit states as a maximally entangled piece plus an orthogonal one.  

\subsection{Evolutions}

A pure state evolution  $|\Psi(t)\rangle$, when local,  assumes the simple form,
\begin{eqnarray}
\Psi (t) = U_1(t)\, \Psi(0)\, U_2^T(t)  \makebox[.5in]{,}  U_1 = e^{-iH_1 t} \makebox[.2in]{,} 
U_2 = e^{ -iH_2t}  \;,
\end{eqnarray}
The density operator, for a two-qudit system in a mixed state, can be conveniently written in the 
$|\bar{A}\rangle$-basis,
\begin{equation}
\hat{\rho}=  \rho_{\bar{A}\bar{B}}  \,  |\bar{A}\rangle\langle \bar{B}| \;..
\end{equation}
Let us initially discuss the Liouville equation for an isolated mixed system,
\begin{eqnarray}
i  \partial_t \hat{\rho} =  [\hat{H}_{\rm S},  \hat{\rho}]  \makebox[.5in]{,} \hat{H}_{\rm S} = \hat{H}_1 + \hat{H}_2  \;,
\label{Liouville}
\end{eqnarray}
and local evolutions (we are using $\hbar =1$). In this case, to compute the right-hand side, we need the commutators, 
\begin{equation}
[\hat{H}_{\rm S}, |\bar{A} \rangle \langle \bar{B}| ] = [\hat{H}_1, |\bar{A} \rangle \langle \bar{B}| ] 
+[\hat{H}_2, |\bar{A} \rangle \langle \bar{B}| ]\;. 
\end{equation}
In particular,
\begin{equation}
 \hat{H}_1 |\bar{A} \rangle  
= T_{\bar{A}} |_{i j} \,   \hat{H}_1    | i j \rangle  = H_1|_{ki}\, T_{\bar{A}} |_{i j} \,    |k j \rangle  = |H_1T_{\bar{A}}\rangle   \;,
 \end{equation}
\begin{equation}
 \hat{H}_2 |\bar{A} \rangle  
= T_{\bar{A}} |_{i j} \,   \hat{H}_2    | i j \rangle  = H_2|_{kj}\, T_{\bar{A}} |_{ij} \,    |ik \rangle  = |T_{\bar{A}} H_2^T \rangle   \;,
 \end{equation}
which imply,
\begin{equation}
\langle\bar{B} |  \hat{H}_1  =  \langle H_1T_{\bar{B}} |  \makebox[.5in]{,} 
\langle\bar{B} | \hat{H}_2 =  \langle T_{\bar{B}} H_2^T|  \;.
 \end{equation}
Therefore, we get,
\begin{eqnarray}
 [\hat{H}_1, |\bar{A} \rangle \langle \bar{B}| ]  &=& |H_1T_{\bar{A}}\rangle\langle T_{\bar{B}}| -|T_{\bar{A}}\rangle \langle H_1 T_{\bar{B}}|  \;,
 \end{eqnarray}
\begin{eqnarray}
 [\hat{H}_2, |\bar{A} \rangle \langle \bar{B}| ]  &=& |T_{\bar{A}}H_2^T \rangle \langle T_{\bar{B}}| -|T_{\bar{A}}\rangle \langle T_{\bar{B}} H_2^T|  \;.
 \end{eqnarray}

In order to solve the Liouville equation, we have to expand the right-hand side of Eq.  (\ref{Liouville}) in terms of the basis $|\bar{A}\rangle\langle\bar{B}|$\,. Similarly,
in a general non-isolated system, the density operator $\hat{\rho}(t)$ satisfies a master equation, that involves commutators with an interaction Hamiltonian that effectively couples the system degrees of freedom with the environment. That is, we must expand states of the form $|T_{\bar{A}} T_{\bar{B}}\rangle$ or $|T_{\bar{A}} T_{\bar{B}}^T\rangle$ in terms of the states $|T_{\bar{C}}\rangle$\,. For this objective, it is convenient to work with the Cartan decompoition of the $\mathfrak{su}(d)$ Lie algebra.

\subsection{Algebraic aspects}
\label{lie}
 
The generators  $T_{A}$, $A=1,2,...,d^2-1$ can be separated into diagonal elements $T_q$, $q=1,\dots, d-1$,  and $d(d-1)$ off-diagonal elements. The latter can be written as combinations,
\begin{equation}
\frac{1}{\sqrt{2}}(E_\alpha + E_{-\alpha})
\makebox[.5in]{,}
\frac{1}{\sqrt{2}i}(E_\alpha - E_{-\alpha})  \;,
\label{Tes}
\end{equation}
where the nonhermitian $E_{\alpha}, E_{-\alpha}$ satisfy,
\begin{equation}
[T_q,T_p]=0 \makebox[.3in]{,}
[T_q,E_{\alpha}]=\vec{\alpha}|_q\, E_{\alpha}
\makebox[.3in]{,}
[E_{\alpha},E_{-\alpha}]= \vec{\alpha}|_q\, T_q\;,
\label{algebrag}
\end{equation}
and $T_q^T=T_q$, $E_\alpha^T = E_ {-\alpha}$. The subindex $\alpha$ indicates a positive $(d-1)$-tuple $\vec{\alpha}$
(positive root) whose $\vec{\alpha}|_q$ component is defined by the previous commutators \cite{footnote}. 

 A weight $\vec{w}$ is defined by the eigenvalues of diagonal generators corresponding to one common eigenvector. 
In the fundamental representation, the diagonal of $T_q$ can be given by,
\begin{equation}
\frac{1}{\sqrt{q(q+1)}}\,(1,\dots,1,-q,0,\dots,0)\;,
\label{diag}
\end{equation}
where the initial $q$ elements are equal to $1$.  The weights of the fundamental representation, $\vec{w}_{i}$, $i=1,\dots,d$, are then given by,
\begin{equation}  
\vec{w}_{i} = (T_1|_{ii}, T_2|_{ii},\dots, T_{d-1}|_{ii}) \;,
\label{wfun}
\end{equation}
and satisfy \cite{28}, 
\begin{equation}
\vec{w}_1+\dots +\vec{w}_d=0\makebox[.3in]{,}
\vec{w}_i\cdot \vec{w}_i = \frac{d-1}{d}
\makebox[.3in]{,}
\vec{w}_i \cdot \vec{w}_j = -\frac{1}{d}
\makebox[.3in]{,}
i\neq j\;.
\label{fundwei}
\end{equation}
The roots can be written as $\vec{\alpha}_{ij} =\vec{w}_i -\vec{w}_j$ ($i\neq j$), with the positive (negative) roots associated with $i<j$ ($i>j$). They satisfy $\alpha^2 = \vec{\alpha} \cdot \vec{\alpha} =2$\,. The root vector $E_\alpha$, for $\vec{\alpha}= \vec{\alpha}_{ij}$ only has a nontrivial element at position $ij$, whose value is $1$\,.

In the Cartan basis, a general two-qudit state is,
\begin{equation}
|\Psi \rangle =  \psi_{\bar q}\, | \bar{q} \rangle+ \psi_{\alpha}\,  |\alpha \rangle \;,
\end{equation} 
where $\bar{q}=0,1, \dots, d-1$, and  $|\alpha \rangle= |E_\alpha\rangle$, with $\alpha$ denoting positive as well as negative roots. The following properties are satisfied, 
\begin{equation}
\langle \bar{q} | \bar{p} \rangle = \delta_{\bar{q}\bar{p}} \makebox[.5in]{,} \langle \bar{q}| \alpha \rangle =0 \makebox[.5in]{,}  \langle \alpha | \beta\rangle =\delta_{\alpha \beta}  \;.
\end{equation}

To compute the evolution equations for the density operator expanded in this basis,
\begin{equation}
\hat{\rho} = \rho_{\bar{q}\bar{p}} \, | \bar{q}\rangle \langle \bar{p} | +  \rho_{\bar{q} \alpha} \, | \bar{q}\rangle \langle \alpha | +  \rho_{\alpha\bar{q}} \, | \alpha\rangle \langle \bar{q} | +
\rho_{\alpha\gamma} \, | \alpha\rangle \langle \gamma  | \;,
\label{Cmixed}
 \end{equation}
we will need the following property,
\begin{equation}
T_{\bar q} T_{\bar{p}}  = g_{\bar{q}\bar{p}\bar{r}}   \,T_{\bar{r}}  \makebox[.5in]{,}  g_{\bar{q}\bar{p}\bar{r}} =  {\rm Tr }\, ( T_{\bar{q}} T_{\bar{p}} T_{\bar{r}} )  \;,
\label{g-prop}
\end{equation} 
where $g_{\bar{q}\bar{p}\bar{r}}$ is symmetric in all its indices. In particular,
\begin{equation}
g_{q00} = 0   \makebox[.5in]{,} g_{qp0} = \frac{\delta_{qp}}{\sqrt{d}}   \;.
\end{equation}
In addition, for a root $\vec{\alpha}=\vec{v}_\alpha -\vec{w}_\alpha$, where $\vec{v}_\alpha$ and $\vec{w}_\alpha$ are fundamental weights,
\begin{equation}
 T_q E_\alpha = \vec{v}_\alpha|_q\, E_\alpha   \makebox[.3in]{,}   E_\alpha T_q = \vec{w}_\alpha|_q\, E_\alpha 
\end{equation}
In this work, states associated with $E_\alpha E_\beta$ will not be needed, as the various Hamiltonians will only involve the system diagonal degrees of freedom. 

It will be useful to denote the elements of the Cartan basis collectively, as $E_{\bar{\mu}}$, where the 
index $\bar{{\mu}}$ takes values on the $d(d-1)$ roots $\vec{\alpha}$, as well as the $d$ numerical values 
$\bar{q}=0,\dots , d-1$, with $E_{\bar{q}} = T_{\bar{q}}$. In this manner, the mixed state (\ref{Cmixed}) can be rewritten as,
\begin{equation}
\hat{\rho} = \rho_{\bar{\mu}\bar{\nu}} \, | E_{\bar{\mu}}\rangle \langle E_{\bar{\nu}} | = \rho_{\bar{\mu}\bar{\nu}} \, | \bar{\mu}\rangle \langle \bar{\nu} | \;.
 \end{equation}

\subsection{Local operations and the fractional (total) phase} 
\label{fev} 

For example, the local unitary evolutions will be taken as, 
\begin{equation}
H_1 = \vec{\beta}_1 \cdot \vec{T}    \makebox[.5in]{,}  H_2 = \vec{\beta}_2 \cdot\vec{T}  \;,
\end{equation}
where the dot product is defined as $\vec{\beta} \cdot \vec{T}= \vec{\beta}|_q\, T_q$\,.
In this respect, we note that a simple operation on an isolated system to generate a total fractional phase at time $\tau=2\pi$  
can be obtained using $\vec{\beta}_1 = \pm \vec{w}$ and $\vec{\beta}_2=0$, or a similar expression interchanging $1 \leftrightarrow 2$\,. In effect, $U(t)= e^{\mp it\,\vec{w} \cdot \vec{T}}  $ satisfies $U(0)=I$
and, 
\begin{equation}
U(2\pi) = e^{ \mp 2\pi i\, \vec{w} \cdot \vec{T}} = e^{\pm i\frac{ 2\pi}{d}}  I
\;.
\label{poss-b} 
\end{equation}

\section{Geometric phase for mixed states } 
\label{GP}

In the kinematic approach \cite{Tong, Tong-err}, the geometric phases for a general mixed state is,
\begin{equation}
\phi_g = \arg \left\{  \sum_{\bar{A}}  \sqrt{\varepsilon_{\bar{A}}(0) \,\varepsilon_{\bar{A}}(\tau) } \, \langle  \Psi_{\bar{A}}(0) | \Psi_{\bar{A}}(\tau) \rangle \, e^{-\int_0^\tau dt\,   \langle  \Psi_{\bar{A}}  | \dot{\Psi}_{\bar{A}}\rangle } \right\}, 
\label{gp}\end{equation}
where $\varepsilon_{\bar{A}}(t)$ and $|\Psi_{\bar{A}}(t) \rangle$ are the eigenvalues and eigenstates of the density operator $\hat{\rho}(t)$, respectively, 
\begin{equation}
\hat{\rho}(t) |\Psi_{\bar{A}}(t) \rangle  = \varepsilon_{\bar{A}}(t)  |\Psi_{\bar{A}}(t) \rangle \;.\end{equation}

If the initial state is pure,  
\begin{equation}
\hat{\rho}(0) = |\Psi(0) \rangle \langle\Psi(0)|\;,
\label{pure-s}
\end{equation}
then the eigenvectors at $t=0$ are given by $|\Psi(0)\rangle$, with eigenvalue $1$, and by an orthogonal degenerate space of eigenvectors, with eigenvalue $0$\,. 
Therefore, in this case, the geometric phase becomes,
\begin{eqnarray}
\phi_g (t) & =& \arg{\langle\Psi(0)|\Psi(t)\rangle} + 
i\int_0^t  ds \,\,\langle\Psi(s)|\dot{\Psi}(s)\rangle \;,
\label{phig}
\end{eqnarray}
which has the same form of that aquired by a pure state, but with an important difference. Here,  $|\Psi(t)\rangle$ is the instantaneous eigenvector of the density operator that at $t=0$ coincides with the initial pure state $|\Psi(0)\rangle$. We shall refer to this particular eigenvector as the ``effective'' state, and will be chosen as the first basis element $|\Psi_0(t)\rangle$, while the two elements associated with initial eigenvalue $0$ will be denoted as $|\Psi_1(t) \rangle$, $|\Psi_2(t) \rangle$.

For future use, we note that for the effective state $|\Psi(t)\rangle$, we can compute the following quantities,
\begin{equation}
I_p=Tr[(\Psi^\dagger \Psi)^{\, p}]=Tr[(\Psi \Psi^\dagger )^{\, p}] \;, 
\end{equation}
$p=1,\dots, d$, which due to the Cayley-Hamilton theorem determine the higher order correlators $I_p$, $p > d$. $I_1$ is simply the 
norm of the eigenvector. If the system were isolated, the remaining correlators would be invariant under local unitary evolutions. In this case, they could be used to divide the total projective space 
$CP^{\,d^2-1}$ into invariant subspaces, representing the well-known fact that entanglement is not affected by such  operations.  For example, $I_2$ would be related to the {\it I-concurrence} \cite{Iconc},
\begin{equation} 
{\cal C}=\sqrt{2(1-I_2)}\;.
\end{equation}
 For a non-isolated system,  which interacts with an environment, it will be interesting  
to analyze how these quantities depend on time.

\section{Master equation} 
\label{master}

Let us now consider a system + environment, with a weak coupling between them. In order to analyze dephasing, we shall consider the following system-environment coupling,
\begin{equation}
\hat{H}_{\rm SE} = \hat{H}_0 + \hat{H}_ {\rm int}
\makebox[.5in]{,}
\hat{H}_0 =\hat{H}_{\rm S}  + \hat{H}_ {\rm E} 
\makebox[.5in]{,}
\hat{H}_{\rm int} =  \hat{\xi}_1^q\, T^1_q + \hat{\xi}_2^q\, T^2_q \;,
\label{couplingxi}  
\end{equation}
where $T^1_q$ ($T^2_q$) acts on the first (second) qudit, and  $\hat{\xi}_1^q$, $\hat{\xi}_2^q$ are the bath degrees of freedom. The evolution of the complete system 
is ruled by the Liouville equation, 
\begin{equation}
\partial_t \hat{\rho}_{\rm SE} =  -i[\hat{H}_{\rm SE}, \hat{\rho}_{\rm SE}] \;,
\end{equation} 
or switching to the interaction picture,
\begin{equation}
\partial_t \hat{\rho}'_{\rm SE} = = -i[\hat{H}'_{\rm int}, \hat{\rho}_{\rm SE}']
\makebox[.5in]{,}
\hat{\rho}_{\rm SE}' =  \hat{U}_0^\dagger\, \hat{\rho}_{\rm SE}\, \hat{U}_0 \;,
\label{eqint}
\end{equation} 
\begin{equation}
\hat{H}'_{\rm int} =  \hat{\xi}_1^q(t)\, T^1_q + \hat{\xi}_2^q(t)\, T^2_q \;,
\end{equation}
where $ \hat{\xi}_1^q(t)$, $ \hat{\xi}_2^q(t)$
represent the free evolution of the bath variables, given by $\hat{H}_{\rm E}$, and we used that $\hat{H}_{\rm S}$ commutes with $\hat{H}_{\rm int}$\,. The environment is supposed to be sufficiently large so as to stay in a stationary state, thus permitting to split the total density operator as,
\begin{equation}
\hat{\rho}_{\rm SE}' \approx \hat{\rho}'(t) \otimes \hat{\rho}_{\rm E}(0) \makebox[.5in]{,}
\hat{\rho}'(t) =  \hat{U}_{\rm S}^\dagger\, \hat{\rho}(t)\, \hat{U}_{\rm S} \;, 
\end{equation}
for all times. It is important to underline that in the Markov regime, we restrict to cases for which the self-correlation functions generated at the environment (due to the coupling interaction) decay faster than typical variation scales in the system.   

The formal solution to Eq. (\ref{eqint}) can be obtained perturbatively using a Dyson expansion. After a series of physical assumptions, the master equation is \cite{BP},
\begin{equation} 
\partial_t \hat{\rho}' = -    \int_0^t ds \, {\mbox Tr}_{\rm E} \left[ \hat{H}'_{\rm int} (t) , \left[ \hat{H}'_{\rm int} (s), 
\hat{\rho}' (t) \times \hat{\rho}_{\rm E}(0)\right] \right] \;.
\end{equation}
To simplify the expressions, we will introduce an index $n=1,2$ to denote the quantities referring to each qudit and will sum over repeated indices. In particular,
\begin{equation}
\hat{H}_{\rm int} =  \hat{\xi}_n^q\, T^n_q \;.
\end{equation}
In this manner, we get,
\begin{eqnarray}
\lefteqn{     \left[ \hat{H}'_{\rm int} (t) , \left[ \hat{H}'_{\rm int} (s), 
\hat{\rho}' (t) \times \hat{\rho}_{\rm E}(0)\right] \right]  } \nonumber \\
&=&  \left[  T^n_q \hat{\xi}_n^q(t) , \left[  T^m_p  \hat{\xi}_m^p(s) , 
\hat{\rho}' (t) \times \hat{\rho}_{\rm E}(0)\right] \right]  \;.
\end{eqnarray}
Now, using that operators acting on the system variables always commute with those acting on the environment, we obtain,
\begin{equation}
[\hat{A}_{\rm S} \hat{A}_{\rm E}, \hat{B}_{\rm S} \hat{B}_{\rm E} ] = \frac{1}{2} \{ \hat{A}_{\rm S}, \hat{B}_{\rm S}  \}  [ \hat{A}_{\rm E}, \hat{B}_{\rm E}  ] +  \frac{1}{2} [ \hat{A}_{\rm S}, \hat{B}_{\rm S} ]  \{ \hat{A}_{\rm E}, \hat{B}_{\rm E}  \} \;.
\end{equation} 
Applying this formula to the double commutator, we get,
\begin{eqnarray}
\lefteqn{     \left[ \hat{H}'_{\rm int} (t) , \left[ \hat{H}'_{\rm int} (s), 
\hat{\rho}' (t) \times \hat{\rho}_{\rm E}(0)\right] \right]  } \nonumber \\
&=&  \frac{1}{4} \Big( [T_q^n, [T_p^m, \hat{\rho}'(t)] ]\, \{ \hat{\xi}_q^n(t) , \{ \hat{\xi}_p^m(s) , \hat{\rho}_{\rm E}(0) \} \} \nonumber \\
&&  +  [T_q^n, \{T_p^m, \hat{\rho}'(t)\} ]\, \{ \hat{\xi}_q^n (t) , [ \hat{\xi}_p^m (s) , \hat{\rho}_{\rm E}(0) ] \}  \nonumber \\ 
&& +  \{ T_q^n, [T_p^m, \hat{\rho}'(t)] \}\, [ \hat{\xi}_q^n(t) , \{ \hat{\xi}_p^m(s) , \hat{\rho}_{\rm E}(0) \} ] \nonumber \\
&& +  \{ T_q^n, \{T_p^m, \hat{\rho}'(t)\} \}\, [ \hat{\xi}_q^n(t) , [ \hat{\xi}_p^m(s) , \hat{\rho}_{\rm E}(0) ] ] \Big)\;.  \nonumber 
\end{eqnarray}
Next, taking the trace over the environment, and using the cyclicity property, the last two terms vanish, obtaining,  
\begin{eqnarray}
\lefteqn{  {\mbox Tr}_{\rm E}  \left[ \hat{H}'_{\rm int} (t) , \left[ \hat{H}'_{\rm int} (s), 
\hat{\rho}' (t) \times \hat{\rho}_{\rm E}(0)\right] \right]  } \nonumber \\
&=&  R_{qp}^{nm}\, [T_q^n, [T_p^m, \hat{\rho}'(t)] ]     + S_{qp}^{nm} \, [T_q^n, \{T_p^m, \hat{\rho}'(t)\} ]\;,
\end{eqnarray}
where we have defined,
\begin{equation}
R_{qp}^{nm}(t) =  \frac{1}{2} \int_0^t ds \, {\mbox Tr}_{\rm E} \left[ \{ \hat{\xi}_q^n(t) ,  \hat{\xi}_p^m(s) \} \hat{\rho}_{\rm E}(0) \right] \;,   \label{noise} 
\end{equation}
\begin{equation}
S_{qp}^{nm}(t) =  \frac{1}{2} \int_0^t ds \, {\mbox Tr}_{\rm E} \left[ [ \hat{\xi}_q^n(t) ,  \hat{\xi}_p^m(s) ] \hat{\rho}_{\rm E}(0) \right] \;.\label{dissp}
\end{equation}
Summarizing, the master equation in the interaction picture is,
\begin{equation}
\partial_t \hat{\rho}' = -  R_{qp}^{nm}\, [T_q^n, [T_p^m, \hat{\rho}'(t)] ]  - S_{qp}^{nm} \, [T_q^n, \{T_p^m, \hat{\rho}'(t)\} ]\;, 
\label{masterEQ}
\end{equation} 
which must be supplemented with,
\begin{equation}
\hat{\rho} (t) = \hat{U}_{\rm S}\, \hat{\rho}'(t)\, \hat{U}_{\rm S} ^\dagger \;, 
\end{equation}
when computing the geometric phase. Note that Eqs. (\ref{noise}) and  (\ref{dissp})  contain noise (decoherence) and dissipation effects, respectively \cite{pra,solidstatedeco}.

By the way, note that the eigenvector $|\Psi(t)\rangle $ of the density operator $\hat{\rho}$ to be used in Eq. (\ref{phig}) can be written as,
\begin{equation}
|\Psi(t)\rangle = \hat{U}_{\rm S}\, |\Psi'(t)\rangle \;,
\end{equation}
where $|\Psi'(t)\rangle $ is the eigenvector of $\hat{\rho}'(t)$\,. Then, solving the master equation in the interaction picture, and obtaining $|\Psi' (t)\rangle $ that tends to the given initial pure state when $t\to 0$, the geometric phase will be given by,
\begin{eqnarray}
\phi_g(t) & =& \arg{\langle\Psi(0)|'\, \hat{U}_{\rm S}(t)\, |\Psi(t)\rangle' } + 
i\int_0^t ds \,\,\langle\Psi(s)|'  \dot{\Psi}(s)\rangle' + \int_0^t ds \,\,\langle\Psi(s)|' \hat{H}_{\rm S} |\Psi(s)\rangle' \;.\nonumber \\  
\label{geo-pure}
\end{eqnarray}

From Eq.(\ref{masterEQ}), the next step is to compute commutators of the form $[ T_q^n , \cdot\,]$\,. For this aim, we calculate the commutator with a general combination of $T_q^n$'s, with arbitrary coefficients $\chi_q^n$, choosing their values  as appropriate at the end. As a term  proportional to the identity operator in the Hamiltonian does not affect the commutators, we are interested in computing them with Cartan generators, acting on different qudits. A long but straight calculation finally leads to, 
\begin{eqnarray} 
[ \chi_1^{q}\, T^1_{q} + \chi_2^{q}\, T^2_{q} ,\hat{\rho}']  =   \rho'_{\bar{{\mu}}\bar{{\nu}}}\left( \chi_1^{q}\, [E^1_{q} , |\bar{{\mu}}\rangle\langle \bar{{\nu}}|] + \chi_2^{q}\, [E^2_{q} , |\bar{{\mu}}\rangle\langle \bar{{\nu}}|]\right) = C_{\bar{\mu} \bar{\nu}} |\bar{\mu} \rangle \langle\bar{\nu}|  \;,
\end{eqnarray}
\begin{eqnarray} 
&& C_{\bar{\mu} \bar{\nu}} |\bar{\mu} \rangle \langle\bar{\nu}|  = g_{q\bar{p}\bar{r}}\, (\chi_1^{q} + \chi_2^{q} )\,( \rho'_{\bar{p}\bar{{\nu}}}\, |T_{\bar{r}}\rangle\langle E_{\bar{{\nu}}}| - \rho'_{\bar{{\mu}}\bar{p}}\, |E_{\bar{{\mu}}}\rangle \langle T_{\bar{r}} |) \nonumber \\
&& +  (\chi_1^{q}\, 
\vec{v}_\alpha|_q + \chi_2^{q}\,\vec{w}_\alpha|_q )\, ( \rho'_{\alpha\bar{{\nu}}}\, |E_\alpha \rangle \langle E_{\bar{{\nu}}}|   - \rho'_{\bar{{\mu}}\alpha}\,  |E_{\bar{{\mu}}}\rangle \langle E_{\alpha}|)  \;,
 \end{eqnarray}
and the coefficients of the commutator in the Cartan basis result,
 \begin{equation}
C_{\bar{r}\bar{s}} = (\chi_1^{q} + \chi_2^{q} )\,(  g_{q\bar{r}\bar{p}}\, \rho'_{\bar{p}\bar{{s}}}-  g_{q\bar{s}\bar{p}}\, \rho'_{\bar{{r}}\bar{p}})
\label{e1}
 \end{equation}
 \begin{equation}
C_{\alpha\bar{r}} =   (\chi_1^{q}\, 
\vec{v}_\alpha|_q + \chi_2^{q}\,\vec{w}_\alpha|_q )\, \rho'_{\alpha\bar{{r}}}  - g_{q\bar{p}\bar{r}}\, (\chi_1^{q} + \chi_2^{q} )\,\rho'_{\alpha\bar{p}}
 \end{equation}
  \begin{equation}
C_{\bar{r}\alpha} = - (\chi_1^{q}\, 
\vec{v}_\alpha|_q + \chi_2^{q}\,\vec{w}_\alpha|_q )\, \rho'_{\bar{{r}}\alpha}+g_{q\bar{p}\bar{r}}\, (\chi_1^{q} + \chi_2^{q} )\, \rho'_{\bar{p} \alpha }
 \end{equation}
  \begin{equation}
C_{\alpha\gamma} =  [\chi_1^{q}\,( 
\vec{v}_\alpha   - \vec{v}_\gamma)|_q + \chi_2^{q}\,(\vec{w}_\alpha - \vec{w}_\gamma)|_q ] \, \rho'_{\alpha\gamma} \;.
\label{e4}
 \end{equation}

\subsection{Some consequences}

Note that the coefficients $C_{\bar{r}\bar{s}} $,  with indices associated with diagonal generators only depend on components of $\rho'$ with this type of indices, and therefore, the same situation applies to $D_{\bar{r}\bar{s}} $\,. A similar situation occurs with coefficientes containing indices associated with either diagonal/off-diagonal, off-diagonal/diagonal, or off-diagonal/off-diagonal generators. 
The master equation in the interaction picture would be,
\begin{equation}
\partial_t \hat{\rho}' (t) =  \partial_t \rho'_{\bar{\mu} \bar{\nu}} |\bar{\mu} \rangle \langle\bar{\nu}| = -D_{\bar{\mu} \bar{\nu}} |\bar{\mu} \rangle \langle\bar{\nu}|  \;,  
\end{equation}
that is,
\begin{equation}
\partial_t \rho'_{\bar{\mu} \bar{\nu}} = - D_{\bar{\mu} \bar{\nu}}   \;.
\end{equation}
As they are linear, if the initial state does not involve off-diagonal degrees $|\alpha \rangle$, that is,
\begin{equation}
\hat{\rho}'(0)  = \rho'_{\bar{q}\bar{p}}(0) \, | \bar{q}\rangle \langle \bar{p} |  \;,
\label{assumed}
 \end{equation}
then, the master equations reduce to solving the $\bar{r}$-sector. Then, in this case we are only interested in emphasizing that,
\begin{equation}
[ T^m_{q'}   ,\hat{\rho}']  =  C_{q'\bar{r}\bar{s}}  \, |\bar{r} \rangle \langle\bar{s}| + \dots \makebox[.5in]{,}    C_{q'\bar{r}\bar{s}}  =  g_{q'\bar{r}\bar{p}}\, \rho'_{\bar{p}\bar{{s}}}-  g_{q'\bar{s}\bar{p}}\, \rho'_{\bar{{r}}\bar{p}}
\, \end{equation}
where the missing kets and bras are associated with at least one off-diagonal generator (and depend on $n$). The same applies to the double commutator, 
\begin{equation}
[ T^n_{q''}   ,  [T_{q'}^m\,, \hat{\rho}'(t)]   ]  =  D_{q''\bar{r}\bar{s}}  \, |\bar{r} \rangle \langle\bar{s}| + \dots \end{equation}
with,
\begin{equation}
D_{q''\bar{r}\bar{s}}  =  g_{q''\bar{r}\bar{p}}\, C_{q'\bar{p}\bar{{s}}}-  g_{q''\bar{s}\bar{p}}\, C_{q'\bar{{r}}\bar{p}}
\makebox[.5in]{,}  C_{q'\bar{p}\bar{s}} =   g_{q'\bar{p}\bar{q}}\, \rho'_{\bar{q}\bar{{s}}}-  g_{q'\bar{s}\bar{q}}\, \rho'_{\bar{{p}}\bar{q}}  \;,
\end{equation}
that is,
\begin{eqnarray}
D_{q''\bar{r}\bar{s}} & = &    g_{q''\bar{r}\bar{p}} \,  g_{q'\bar{p}\bar{q}}\, \rho'_{\bar{q}\bar{{s}}}- g_{q''\bar{r}\bar{p}} \,g_{q'\bar{s}\bar{q}}\, \rho'_{\bar{{p}}\bar{q}} -      g_{q'\bar{r}\bar{q}}\, \rho'_{\bar{q}\bar{{p}}}\, g_{q''\bar{s}\bar{p}} + g_{q''\bar{s}\bar{p}}\, g_{q'\bar{p}\bar{q}}\, \rho'_{\bar{{r}}\bar{q}}  \nonumber \\  
 & = &    g_{q''\bar{r}\bar{p}} \,  g_{q'\bar{p}\bar{q}}\, \rho'_{\bar{q}\bar{{s}}}-  g_{q''\bar{r}\bar{p}} \,  \rho'_{\bar{{p}}\bar{q}} \, g_{q'\bar{q}\bar{s}} -
 g_{q'\bar{r}\bar{q}}\, \rho'_{\bar{q}\bar{{p}}}\, g_{q''     \bar{p}    \bar{s}  }+\rho'_{\bar{{r}}\bar{q}}  \,  g_{q'   \bar{q}  \bar{p}  }\, g_{q''\bar{p}\bar{s}} \;.\nonumber \\
\label{Dcoeff}
\end{eqnarray}
We shall assume that the $S$-coefficients can be disregarded. This is a good approximation for a thermal ohmic-environment composed by an infinite set of quantum harmonic oscillators, in the limit of high temperatures. Then, the master equation for the diagonal-diagonal components is,
\begin{equation}
\partial_t \rho'_{\bar{r}\bar{s}}  =-  R_{q''q'} (    g_{q''\bar{r}\bar{p}} \,  g_{q'\bar{p}\bar{q}}\, \rho'_{\bar{q}\bar{{s}}}-  g_{q''\bar{r}\bar{p}} \,  \rho'_{\bar{{p}}\bar{q}} \, g_{q'\bar{q}\bar{s}} -
 g_{q'\bar{r}\bar{q}}\, \rho'_{\bar{q}\bar{{p}}}\, g_{q''     \bar{p}    \bar{s}  }+\rho'_{\bar{{r}}\bar{q}}  \,  g_{q'   \bar{q}  \bar{p}  }\, g_{q''\bar{p}\bar{s}}   ) \;, 
\end{equation} 
where we defined,
\begin{equation}
R_{q''q'}  =  R_{q''q'}^{11} +   R_{q''q'}^{12} + R_{q''q'}^{21} + R_{q''q'}^{22} \;.
\end{equation}
These quantities are symmetric under the interchange $q'\leftrightarrow q''$\,.
Then, in our case, the master equation in the interaction picture can be written in terms of a reduced ($d\times d$) density matrix $\tilde{\rho}$, with elements $\rho'_{\bar{r}s\bar{}}$ as
\begin{equation}
\partial_t \tilde{\rho} =   - R_{q''q'}  \,  [G_{q''}, [G_{q'},\tilde{\rho}]] \;,
\label{red-mas}
\end{equation}
where $G_q$ is the $d\times d$ matrix with elements $g_{q\bar{r}s\bar{}}$\,.  Some of the componentes are,
\begin{equation}
g_{q00} = 0 \makebox[.5in]{,} g_{q0p} = \frac{ \delta_{qp} }{\sqrt{d}} \;.
\label{alr} 
\end{equation}

\subsection{$SU(2)$}

We note that for two-qubit systems ($d=2$), the index $q$ can only be $1$, so that,
\begin{equation}
\partial_t \tilde{\rho} =   - R_{11}  \,  [G_{1}, [G_{1},\tilde{\rho}]] \;,
\end{equation}
and $ g_{111} = {\rm Tr}\, (T_1 T_1 T_1)=0$, as $T_1$ is proportional to a Pauli matrix,
\[  T_1 = \frac{1}{\sqrt{2}} \left( \begin{array}{cc}
1 & 0 \\
0 & -1  
 \end{array} \right) \;.\] 
Then,  Eq.  (\ref{alr}) implies,
\[  G_1 = \frac{1}{\sqrt{2}} \left( \begin{array}{cc}
0 & 1\\
1 & 0  
 \end{array} \right) \;,\] 
and  the coefficients in the master equation, in the Cartan basis, reduce to,
\begin{eqnarray}
D_{00 }  & = &   R_{11}  \,  ( \rho'_{00} - \rho'_{11} )  \;.
\end{eqnarray}
\begin{eqnarray}
D_{01} 
& = &   R_{11}\, (\rho'_{0 1 }  -\rho'_{1 0} )      \;, 
\end{eqnarray}
\begin{eqnarray}
D_{10}
& = &    R_{11}\, (\rho'_{10 }  -\rho'_{ 01} )           \;, 
\end{eqnarray}
\begin{eqnarray}
D_{11}  & = &     R_{11}\, (\rho'_{11} - \rho'_{00} )
\;.
\end{eqnarray} 

For example, if the initial density matrix is diagonal in the Cartan basis, then it will continue to be diagonal. In this case, the relevant equations are,
\begin{equation}
\partial_t \rho'_{00} =R_{11}\,  ( \rho'_{11} - \rho'_{00})
 \makebox[.5in]{,}
\partial_t \rho'_{11} =R_{11}\,  ( \rho'_{00} - \rho'_{11} ) \;,
 \end{equation}
so that $\rho'_{00}+ \rho'_{11}$ is conserved and,
\begin{equation}
\rho'_{00}(t)  = \frac{ \rho'_{00}(0)}{2} \left(1+ e^{-2R_{11}\, t}   \right) + \frac{ \rho'_{11}(0)}{2} \left(1 -e^{-2R_{11}\, t}   \right) 
 \end{equation}
\begin{equation}
\rho'_{11}(t)  = \frac{ \rho'_{11}(0)}{2} \left(1+ e^{-2R_{11}\, t}    \right) 
+ \frac{ \rho'_{00}(0)}{2} \left(1 -e^{-2R_{11}\, t}    \right) \;.
 \end{equation}
In particular, if at $t=0$ we have the pure maximally entangled state $\rho'_{00}(0)=1$, $\rho'_{11}(0)=0$, the instantaneous eigenvector needed to compute the geometric phase 
is the maximally entangled state $|0 \rangle $, $\phi_g$ will not be affected by decoherence, and will coincide with the fractional phase $\pi$. This is one of the main results in Ref. \cite{bipartite}. For $SU(d)$, due to the nontrivial  coefficients $g_{qp}$, a careful analysis is required. 

\section{The master equation for two-qutrit states}
\label{master2}

For $SU(3)$,  Eq. (\ref{diag}) gives the diagonal Gell-mann matrices, 
\begin{equation} T_1 = \frac{1}{\sqrt{2}} \left( \begin{array}{ccc}
1 & 0 & 0 \\
0 & -1 & 0 \\
0 & 0 & 0 \end{array} \right)  \makebox[.5in]{,} 
T_2 = \frac{1}{\sqrt{6}} \left( \begin{array}{ccc}
1 & 0 & 0 \\
0 & 1 & 0 \\
0 & 0 & -2 \end{array} \right) \;.
\label{T12}
\end{equation}
The explicit form of the Eq. (\ref{red-mas}) is obtained from Eq. (\ref{alr}) together with,  
\begin{equation} 
 {\rm Tr }\, ( T_1 T_1 T_1 )  =0  \makebox[.5in]{,}    {\rm Tr }\, ( T_2 T_1 T_1 )  = \frac{1}{\sqrt{6}}  
\end{equation}
\begin{equation}
    {\rm Tr }\, ( T_1 T_1 T_2 )  =  \frac{1 }{\sqrt{6}}     \makebox[.5in]{,}   {\rm Tr }\, ( T_2 T_1 T_2 ) = 0 
\end{equation}
\begin{equation}
   {\rm Tr }\, ( T_1 T_2 T_2 )  = 0  \makebox[.5in]{,}    {\rm Tr }\, ( T_2 T_2 T_2 )  = - \frac{1 }{\sqrt{6}}  \;.
\end{equation}
Now we have all the elements needed to write the relevant part of the master equation
for,
\[  \tilde{\rho} = \left(    
\begin{array}{ccc}
\rho'_{00} & \rho'_{01} & \rho'_{02} \\ 
\rho'_{10} & \rho'_{11} & \rho'_{12} \\
\rho'_{20} & \rho'_{21} & \rho'_{22} \end{array} \right)  
\;.\] 
It is given by,
\begin{equation}
\partial_t \tilde{\rho} =   - R_{11}  \,  [G_{1}, [G_{1},\tilde{\rho}]]   - R_{12}  \,  [G_{1}, [G_{2},\tilde{\rho}]] -R_{21}  \,  [G_{2}, [G_{1},\tilde{\rho}]] -  R_{22}  \,  [G_{2}, [G_{2},\tilde{\rho}]]  \;,
\label{mesu3}
\end{equation}
with,
\[ G_1 = \left(  \renewcommand*{\arraystretch}{1.2}
\begin{array}{ccc}  
 0 &\frac{ 1}{\sqrt{3}} & 0 \\   
\frac{1}{\sqrt{3}} & 0 & \frac{1}{\sqrt{6}}  \\
0& \frac{1}{\sqrt{6}}  & 0   \end{array}    \right) \makebox[.5in]{,} 
G_2 = \left(  \renewcommand*{\arraystretch}{1.2}
\begin{array}{ccc}  
 0 &  0 & \frac{1}{\sqrt{3}} \\   
0 & \frac{1}{\sqrt{6}} &0  \\
\frac{1}{\sqrt{3}} & 0  & - \frac{1}{\sqrt{6}}    \end{array}    \right) \;.\]

\subsection{Analysis}
\label{analy}

Let us suppose possible couplings in Eq. (\ref{couplingxi}) with suppressed dissipation coefficients. 
If the couplings are of the form,
\begin{equation}
 \hat{\xi}_1^q = \vec{\gamma}_1|^q\, \hat{\xi}   \makebox[.5in]{,} \, \hat{\xi}_2^q =  \vec{\gamma}_2|^q\, \hat{\xi} \;,
 \label{sime}
\end{equation}  
where $\hat{\xi}$ represents a collection of harmonic-oscillator degrees of freedom, that is, 
$\hat{\xi}_q^n(t) =\vec{\gamma}_n|^q\, \hat{\xi}   $\,, we have,
\begin{equation}
R_{qp}^{nm}(t) =  \frac{1}{2}\, \vec{\gamma}_n|^q\, \vec{\gamma}_m|^p\, f(t )  \makebox[.5in]{,} 
f(t) = \int_0^t ds \, {\mbox Tr}_{\rm E} \left[ \{ \hat{\xi}(t) ,  \hat{\xi}(s) \} \hat{\rho}_{\rm E}(0) \right] \;,    
\end{equation}
which implies,
\begin{equation}
R_{qp}  =  \frac{1}{2}\, \left( \vec{\gamma}_1|^q\, \vec{\gamma}_1|^p + \vec{\gamma}_1|^q\, \vec{\gamma}_2|^p+ \vec{\gamma}_2|^q\, \vec{\gamma}_1|^p + \vec{\gamma}_2|^q\, \vec{\gamma}_2|^p \right) f(t )
  \;,
  \label{erres}
\end{equation}
 \begin{equation}
 \vec{\gamma}_1 = (a_1 , b_1) \makebox[.5in]{,}
   \vec{\gamma}_2 = (a_2 , b_2) \;,
 \end{equation} 
 \begin{eqnarray}
R_{11} & = & \frac{1}{2}\, (a_1 + a_2)^2  \, f(t) \nonumber \\
R_{12} & = & \frac{1}{2}\, ( a_1 +  a_ 2 )(b_1 +  b_2 )\, f(t ) \nonumber \\
R_{21} & = & \frac{1}{2}\, ( a_1 +  a_ 2 )(b_1 +  b_2 )\, f(t ) \nonumber \\
R_{22} & = &  \frac{1}{2}\, (b_1 + b_2)^2 \,  f(t )   \;.
\end{eqnarray}
When both qudits are coupled to the environment in the same way, we have,
\begin{equation}
\vec{\gamma}_1 = \vec{\gamma}_2 = \vec{\gamma}=(a,b)
\label{gammas}
\end{equation}
\begin{equation}
R_{11} = 2a^2\, f(t) \makebox[.5in]{,} R_{22} = 2b^2\, f(t)
\makebox[.5in]{,} 
R_{12}  = R_{21} =  2 ab\, f(t ) \;.
\end{equation}
In this case, the master equation (\ref{mesu3}) reads,
\[
\partial_t \tilde{\rho} =   - 2f(t) \left[ a^2  \,  [G_{1}, [G_{1},\tilde{\rho}]]   + ab  \,  [G_{1}, [G_{2},\tilde{\rho}]] + ab  \,  [G_{2}, [G_{1},\tilde{\rho}]] +  b^2  \,  [G_{2}, [G_{2},\tilde{\rho}]] \right]   \;,
\] 
and by defining,
\begin{equation}
\zeta_1 = \sqrt{2f(t)}\, a \makebox[.5in]{,} \zeta_2 = \sqrt{2f(t)}\, b \;,
\label{zetas}
\end{equation}
\begin{equation}  
G  =   \zeta_1\, G_{1} + \zeta_ 2\, G_{2} = \left(  \renewcommand*{\arraystretch}{1.2}
\begin{array}{ccc}  
 0 &\frac{ \zeta_1}{\sqrt{3}} & \frac{\zeta_2}{\sqrt{3}} \\   
\frac{\zeta_1}{\sqrt{3}} & \frac{\zeta_2}{\sqrt{6}} & \frac{\zeta_1}{\sqrt{6}}  \\
\frac{\zeta_2}{\sqrt{3}} & \frac{\zeta_1}{\sqrt{6}}  & - \frac{\zeta_2}{\sqrt{6}}    \end{array} \right)  \;,
\end{equation} 
it can be rewritten as,
\begin{equation}
\partial_t \tilde{\rho} =  -[G, [G,\tilde{\rho}]] \;.
\end{equation}

Using the eigenvalues and eigenvectors of $G$, it is possible to write $G = R D R^{-1}$ where D is diagonal.  In fact, one finds that $R$ does not depend on $\zeta_1$, $\zeta_2$ ($R R^T = R^T R = I$):
\[
R=  \left( \begin{array}{ccc}
-\frac{1}{\sqrt{3}} & \frac{1}{\sqrt{3}} & \frac{1}{\sqrt{3}}  \\
0 & -\frac{1}{\sqrt{2}}  & \frac{1}{\sqrt{2}}  \\
\frac{\sqrt{2}}{\sqrt{3}}  & \frac{1}{\sqrt{6}}  & \frac{1}{\sqrt{6}}  \end{array} \right),
\]
 and 
\[  D = \left( 
\begin{array}{ccc}  
 -\sqrt{\frac{2}{3}}\zeta_2 & 0 & 0 \\   
0 & \frac{-3\sqrt{2}\zeta_1 + \sqrt{6}\zeta_2}{6} & 0  \\
0  & 0 &  \frac{3\sqrt{2}\zeta_1 + \sqrt{6}\zeta_2}{6}   \end{array} \right)..
\] 
This is useful as we can propose a solution of the form,
\begin{equation}
\tilde{\rho}(t) = R\, \sigma (t) R^{-1}
\makebox[.5in]{,} 
\partial_t \sigma =  - [D, [D,\sigma ]] \;.
\label{newmastereq}
\end{equation}
Note that as $R$ is time-independent, the equation for $\sigma$ is valid even for time-dependent $\zeta_1, \zeta_2$.
 Solving this equation we get
 \begin{equation} \sigma(t) = \left(  
\begin{array}{ccc}  
\sigma_{00}(0)  &  e^{-\frac{t}{2} (\zeta_1 - \sqrt{3}\zeta_2)^2} \sigma_{01}(0) &   e^{-\frac{t}{2} (\zeta_1 + \sqrt{3}\zeta_2)^2} \sigma_{02}(0)\\   
 e^{-\frac{t}{2} (\zeta_1 - \sqrt{3}\zeta_2)^2} \sigma_{10}(0) & \sigma_{11}(0)  & e^{- 2 t\, \zeta_1^2} \sigma_{12}(0) \\
e^{-\frac{t}{2} (\zeta_1 + \sqrt{3}\zeta_2)^2} \sigma_{20}(0) & e^{- 2 t\, \zeta_1^2} \sigma_{21}(0)  & \sigma_{22}(0) \end{array} \right)   \;,   
\label{sigmat}
\end{equation} 
where we assumed $f(t)$ constant.

We recall that the decoupling of the off-diagonal sector is due to the choice of initial density matrix (cf. Eq. (\ref{assumed})), as well as a free and interaction dynamics based on diagonal degrees. Another simplification was done to arrive at Eq. (\ref{geo-pure}), where an initial pure state was assumed. Both considerations lead to the initial condition, 
\begin{equation}
\hat{\rho}'(0)  = |\Psi(0) \rangle \langle \Psi(0)| \makebox[.5in]{,}  | \Psi(0) \rangle
= a_{\bar{q}}\, |\bar{q} \rangle \makebox[.5in]{,} a_{\bar{q}} \bar{a}_{\bar{q}} =1\;,
 \end{equation}
\begin{equation}
\hat{\rho}'(0)    =  \rho'_{\bar{q}\bar{p}}(0) \, | \bar{q}\rangle \langle \bar{p} |  = a_{\bar{q}}\,  \bar{a}_{\bar{p}}\,  |\bar{q} \rangle  \langle \bar{p} |  \;,
 \end{equation}
 that is,
 \[
  \tilde{\rho}(0) =  \left( 
\begin{array}{ccc}  
 a_0  \\   
 a_1   \\
 a_2    \end{array} \right)   \Big( 
  \bar{a}_0 \hspace{.2cm}   \bar{a}_1 \hspace{.2cm}     \bar{a}_2    \Big) =
 \left(  
\begin{array}{ccc}  
 a_0  \bar{a}_0 &  a_0  \bar{a}_1 &  a_0  \bar{a}_2 \\   
 a_1  \bar{a}_0 &  a_1  \bar{a}_1 &  a_1  \bar{a}_2  \\
 a_2  \bar{a}_0  &  a_2  \bar{a}_1 &   a_2  \bar{a}_2   \end{array} \right)
 \]
\[
 \sigma(0) =  \left( 
\begin{array}{ccc}  
 b_0  \\   
 b_1   \\
 b_2    \end{array} \right)   \Big( 
  \bar{b}_0 \hspace{.2cm}   \bar{b}_1 \hspace{.2cm}     \bar{b}_2    \Big) \makebox[.5in]{,}  \left( 
\begin{array}{ccc}  
 b_0  \\   
 b_1   \\
 b_2    \end{array} \right)  = R^T  \left( 
\begin{array}{ccc}  
 a_0  \\   
 a_1   \\
 a_2    \end{array} \right) \;,  
\]
\[
b_0 =  \frac{\sqrt{2}\, a_2-a_0}{\sqrt{3}} \makebox[.3in]{,} b_1= \frac{a_2 -\sqrt{3}\, a_1 + \sqrt{2}\, a_0}{\sqrt{6}}   \makebox[.3in]{,}  b_2=   \frac{a_2 +\sqrt{3}\, a_1 + \sqrt{2}\, a_0}{\sqrt{6}} \;.
\]
Now,  we need the eigenvalues of $\hat{\rho}'(t)$ and the corresponding eigenvectors. At $t=0$, as the state is pure, there is one eigenvalue  equal to $1$, whose eigenvector is $|\Psi(0) \rangle $, and the remaining eigenvalues are $0$. In order to evaluate the geometric phase, we shall look for the eigenvector,
\begin{equation}
\hat{\rho}'(t)|\Psi'  (t) \rangle = \epsilon(t) | \Psi'  (t) \rangle  \makebox[.5in]{,}
 | \Psi'  (t) \rangle
= u_{\bar{q}}(t)\, |\bar{q} \rangle  \makebox[.5in]{,} u_{\bar{q}} \bar{u}_{\bar{q}} =1\;,
\label{comb-s} 
\end{equation}
or in matrix notation,
\begin{equation} 
 \tilde{\rho}(t) \,  u(t) = \varepsilon(t)\, u(t)
\makebox[.5in]{,} u(t) =\left( 
\begin{array}{ccc}  
 u_0(t)  \\   
u_1(t)\\
u_2(t)\end{array} \right)  \;,
\label{m-not}
 \end{equation}
with the initial condition,
\[
\varepsilon(0)=1  \makebox[.5in]{,} u_{\bar{q}}(0) =a_{\bar{q}} \;. 
\]
In what follows, this particular eigenvector will be referred to as the ``effective'' state.
We can also write $u(t)$ in terms of the corresponding eigenvector of $\sigma(t)$,

\[
u(t) = R\, v(t) \makebox[.5in]{,}  \sigma(t) \,  v(t) = \epsilon(t)\, v(t) \;.
\] 
The effective state is associated with the matrix (cf. Eq. (\ref{comb-s})),

\begin{eqnarray} 
 \Psi'  (t)  &=&   \frac{u_0(t)}{\sqrt{3}} \, I + u_1(t) \, T_1 + u_2(t) \,T_2 \;,
 \label{pprima}
 \end{eqnarray}
with the matrices $T_1$, $T_2$ given in Eq. (\ref{T12})\,.

Now, we have all the ingredients needed to compute different observables. For example, operating on the first qutrit of the system, that is, $\vec{\beta}_1 = \vec{w}$ and $\vec{\beta}_2=0$, the geometric phase generated at time $t$ is,

\begin{equation}
\phi_g(t) = \arg \, {\rm Tr}  \left[ (\Psi'(0))^\dagger \,  U_{\rm S}(t)\, \Psi'(t) \right]
+ 
i\int_0^t ds\, ( u_0^\ast \, \dot{u}_0 + u_1^\ast \, \dot{u}_1 + u_2^\ast \, \dot{u}_2 )
+ \int_0^t ds \,\, {\rm Tr}  \left[ (\Psi'(s))^\dagger  H_{\rm S} \,\Psi'(s) \right] \;.
\end{equation}
\begin{equation}
H_{\rm S} = \vec{w} \cdot \vec{T} \makebox[.5in]{,} U_{\rm S}(t)= e^{- it\,\vec{w} \cdot \vec{T}}  \;.
\label{Hese}
\end{equation}

If the system were isolated, to generate fractional phases, $\vec{w}$ must be a fundamental weight. For instance, from Eq. (\ref{T12}) we can take for $\vec{w}$,

\begin{equation}
\vec{w}_1 =\left(0 , \frac{-2}{\sqrt{6}} \right)  
\makebox[.5in]{,} 
\vec{w}_2 = \left(- \frac{1}{\sqrt{2}} , \frac{1}{\sqrt{6}} \right) 
\makebox[.5in]{,} 
\vec{w}_3 = \left( \frac{1}{\sqrt{2}} , \frac{1}{\sqrt{6}} \right)  \;.
 \end{equation} 
For example, if the first tuple is used, 

\begin{equation}
H_{\rm S} =   \frac{-2}{\sqrt{6}} \, T_2 = \frac{1}{3} \left( \begin{array}{ccc}
-1 & 0 & 0 \\
0 & -1 & 0 \\
0 & 0 & +2 \end{array} \right) \makebox[.5in]{,} U_{\rm S}(t)=  \left( \begin{array}{ccc}
e^{\frac{1}{3}it} & 0 & 0 \\
0 & e^{\frac{1}{3} it} & 0 \\
0 & 0 & e^{-\frac{2}{3} it}\end{array} \right)  \;. \nonumber 
\end{equation}
Note that after a period of  the free evolution $\tau = 2\pi$, we have $U_{\rm S}(2\pi) = e^{\frac{2\pi}{3}i}\, I$.
It is also interesting to define and analyze an effective concurrence that in general will be time-dependent,

\begin{equation}
{\cal C}(t)=\sqrt{2(1-{\rm  Tr}\,[(\Psi \Psi^\dagger )^{2}]   )} =\sqrt{2(1-{\rm  Tr}\,[(\Psi' \Psi'^\dagger )^{2}]   )}   \;.
\end{equation}
where we used $\Psi(t) = U_{\rm S} (t) \Psi'(t) $, and the unitarity of the free evolution. This is a quantity that at $t=0$ coincides with the concurrence $C_0$ of the initial pure state \cite{Iconc,concurrence}, and characterizes the instantaneous eigenvector needed to compute the geometric phase. In this respect, if the initial state is characterized by real amplitudes $a_q$, the normalized eigenvector will have real coefficients $u_q$ and the geometric phase reduces to,
\begin{eqnarray}
\phi_g(t) & =& \arg \, {\rm Tr}  \left[ (\Psi'(0))^\dagger \,  U_{\rm S}(t)\, \Psi'(t) \right]
 + \int_0^t ds \,\, {\rm Tr}  \left[ (\Psi'(s))^\dagger  H_{\rm S} \,\Psi'(s) \right]  \;.
 \label{phasep}
\end{eqnarray}
This looks like the Simon-Mukunda geometric phase for an isolated system \cite{Simon-Mukunda}, but computed for an effective state with time-dependent concurrence.

\section{The effective concurrence $C(t)$}
\label{concurrence}

\begin{figure}[h]
\centering   
	\subfloat[$ 0.90$]{\includegraphics[width=.31\textwidth]{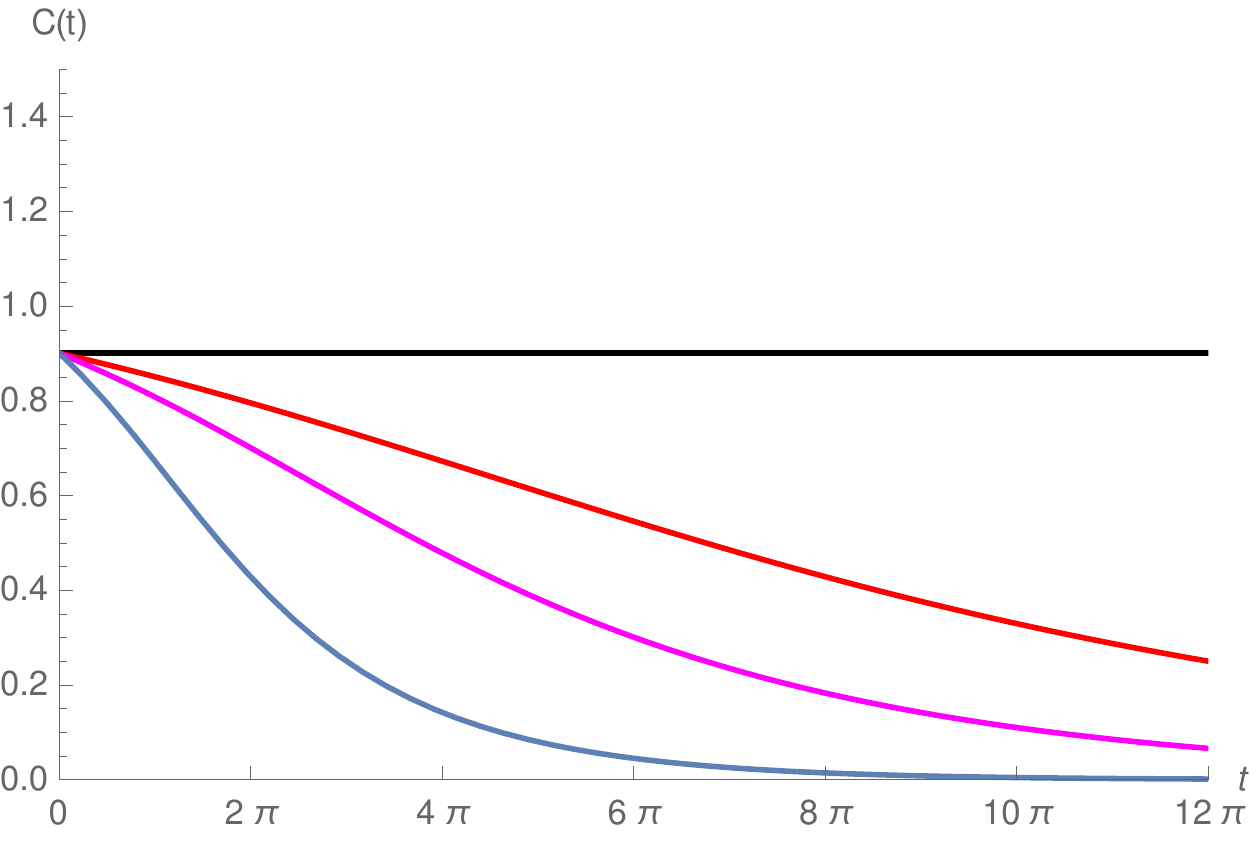}} \hspace{.2cm}  
	\subfloat[$ 0.99$]{\includegraphics[width=.31\textwidth]{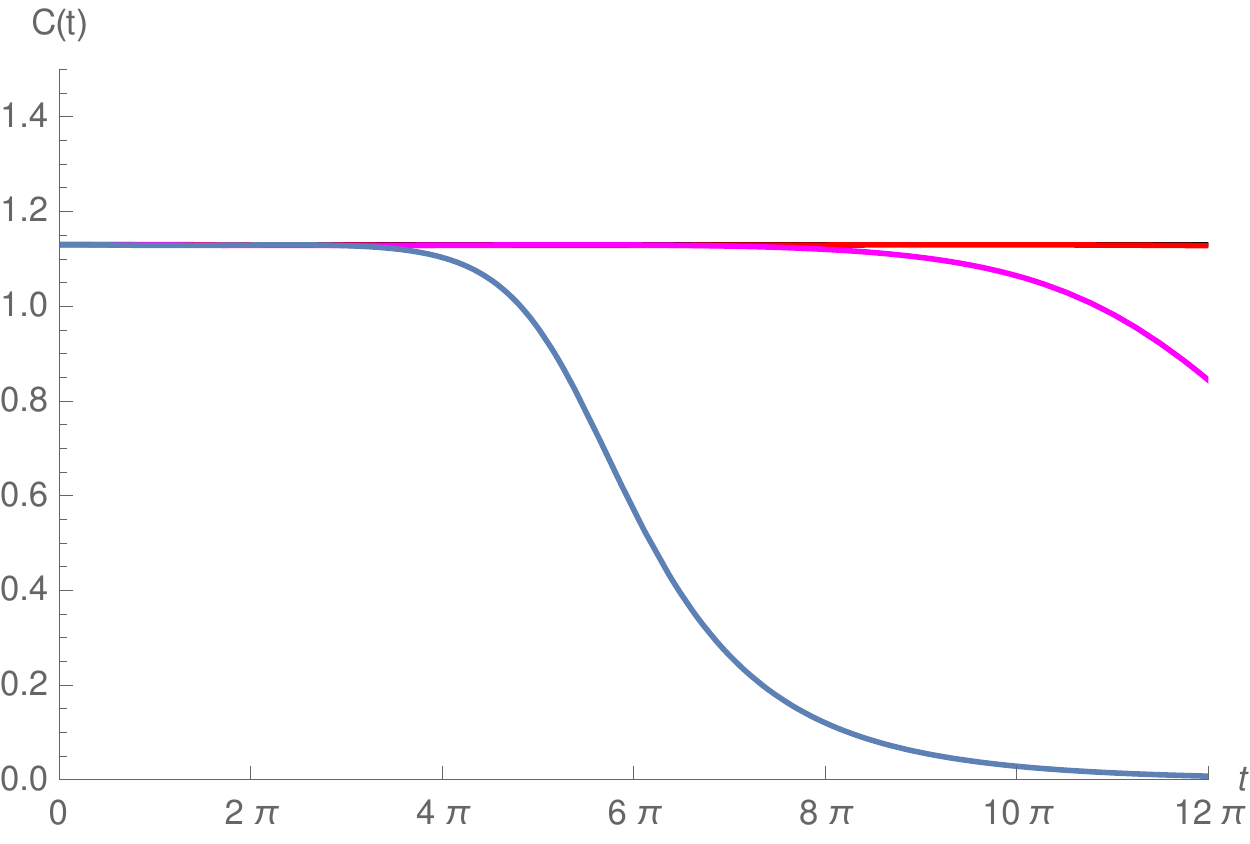}} \hspace{.2cm} 
\subfloat[0.993 ]{\includegraphics[width=.31\textwidth]{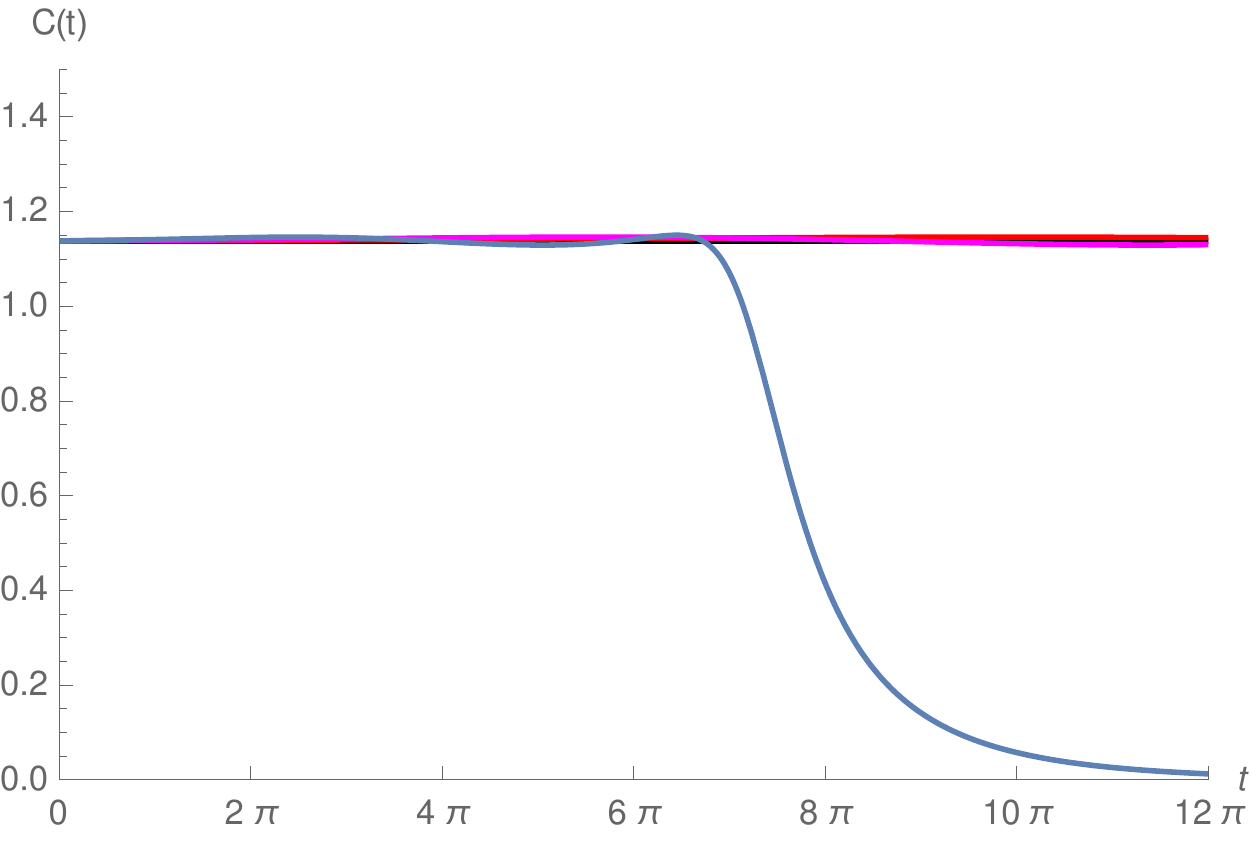}} 
\caption{The effective concurrence $C(t)$.  The plots are labeled with the value of $a_0$ considered.  Different couplings $(\zeta_1, \zeta_2)$ are shown as: $(0 , 0)$ in black, $(0.15 , 0.15)$ in red, $(0.2 , 0.2)$ in magenta, and $(0.3 , 0.3)$ in blue.  }   
\label{conj1}   
\end{figure}

Let us consider a rather general case where $\zeta_1$ and $\zeta_2$ are such that all the off-diagonal components in Eq. (\ref{sigmat}) are damped, and the diagonal components of $\sigma$ are nonzero. That is, for $t> t_{\rm A}$ (a characteristic asymptotic time), we have,
 \begin{equation} \sigma(t) \to \sigma_{\rm A} = \left(   
\begin{array}{ccc}  
\sigma_{00}(0)  &  0 &  0\\   
0 & \sigma_{11}(0)  & 0 \\
0 & 0  & \sigma_{22}(0) \end{array} \right)   \;,   
\end{equation}  
and the asymptotic eigenvalues of $\tilde{\rho}(t)$ are,
 \[  
\sigma_{00}(0)  =  \frac{|\sqrt{2}\, a_2-a_0|^2}{3} \makebox[.3in]{,} \sigma_{11}(0)= \frac{|a_2 -\sqrt{3}\, a_1 + \sqrt{2}\, a_0|^2}{6}   \makebox[.3in]{,}  \sigma_{22}(0)=  \frac{|a_2 +\sqrt{3}\, a_1 + \sqrt{2}\, a_0|^2}{6} \;.
\]
Now, for initial states such that $\sigma_{00}(0)$, $\sigma_{11}(0)$ and $\sigma_{22}(0)$ are all different,
the instantaneous eigenvectors of $\tilde{\rho}(t)$ will necessarily evolve to get the asymptotic values,
\[
R\left(\begin{array}{ccc}  
 1  \\   
0\\
0 \end{array} \right) =
\left( \begin{array}{ccc}
-\frac{1}{\sqrt{3}}  \\
0  \\
\frac{\sqrt{2}}{\sqrt{3}}    \end{array} \right)
 \makebox[.5in]{,} 
 R\left(\begin{array}{ccc}  
 0  \\   
1\\
0 \end{array} \right) =  \left( \begin{array}{ccc}
 \frac{1}{\sqrt{3}}   \\
 -\frac{1}{\sqrt{2}}  \\
 \frac{1}{\sqrt{6}}  \end{array} \right)
 \makebox[.5in]{,}
 R\left(\begin{array}{ccc}  
 0  \\   
0\\
1 \end{array} \right) = \left( \begin{array}{ccc}
 \frac{1}{\sqrt{3}}  \\
\frac{1}{\sqrt{2}}  \\
 \frac{1}{\sqrt{6}}  \end{array} \right) \;.
\] 
In particular, the instantaneous eigenvector $u(t)$ must tend to one of them, and the asymptotic behavior of $\Psi'(t) \to \Psi'_A$ (cf Eq. (\ref{pprima})), needed to compute the geometric phase, must be one of the following,
 \begin{equation} 
\left(  
\begin{array}{ccc}  
0  &  0 &  0\\   
0 & 0  & 0 \\
0 & 0  & -1 \end{array} \right) 
 \makebox[.5in]{,} 
 \left(  
\begin{array}{ccc}  
0  &  0 &  0\\   
0 & 1  & 0 \\ 
0 & 0  & 0 \end{array} \right) 
 \makebox[.5in]{,} 
\left(  
\begin{array}{ccc}  
1  &  0 &  0\\   
0 & 0  & 0 \\
0 & 0  & 0 \end{array} \right) 
   \;. 
\end{equation}   
Then, we can see that the asymptotic effective concurrence $C(t)$ will tend to zero. In Fig.  
\ref{conj1}, we show the numerical results for $C(t)$ (the numerical calculations in this work were performed with {\sc Mathematica}), using an initial state characterized by
\begin{equation}
a_1= a_2 = \frac{\sqrt{1- a_0^2}}{\sqrt{2}} \;.
\end{equation}
The time scale in multiples of $2\pi$ is chosen to compare with typical times for the isolated system, which has a period of $2\pi$. The curves correspond to different couplings $(\zeta_1, \zeta_2)$: $(0 , 0)$ in black, $(0.15 , 0.15)$ in red, $(0.2 , 0.2)$ in magenta, and $(0.3 , 0.3)$ in blue, a convention that will be used in all the figures. For $a_0 =0.90$, $C_0 \approx 0.9$ (a), we clearly see the decay of  
$C(t)$ to zero, when the system is coupled to the environment. An interesting behavior arises  as the initial state gets closer to the MES state. At $a_0=0.99$, $C_0 \approx 1.13$ (b) we observe that the effective concurrence is more protected against the effects of the environment. A little bit further, at  $a_0=0.993$ (c), and for $(\zeta_1,\zeta_2)$ given by  $(0 , 0)$, $(0.15 , 0.15)$, or $(0.2 , 0.2)$, the curves $C(t)$ cannot be distinguished in the scales displayed in the plot (up to $t=12 \pi$). They are stabilized at $C(0)$, eventually decaying to zero at larger times. For  $(\zeta_1,\zeta_2)= (0.3 , 0.3)$, $C(t)$ is stabilized up to a time between $6\pi$ and $8\pi$, where it starts decaying to zero.

\begin{figure}[h] 
\centering 
	\subfloat[0.994]{\includegraphics[width=.31\textwidth]{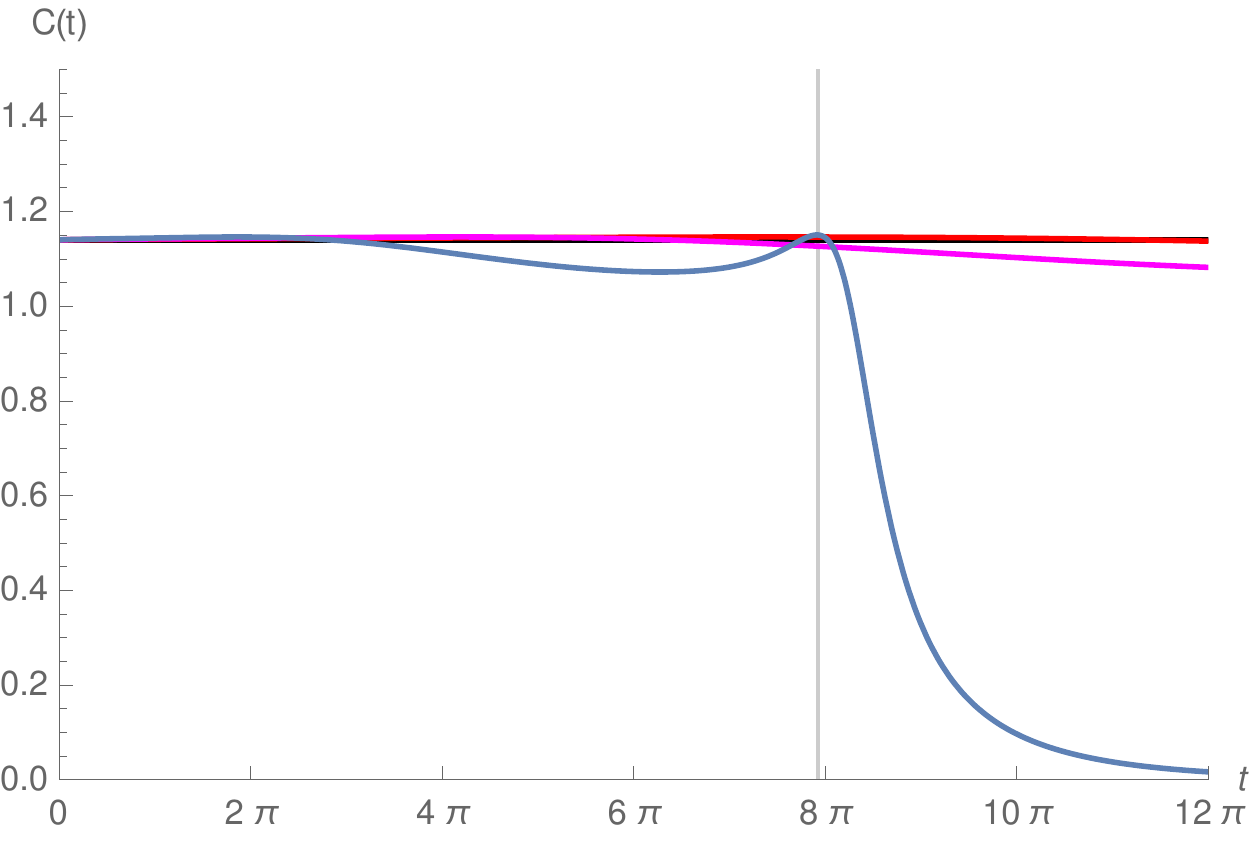}}  \hspace{.2cm} 
	\subfloat[0.995]{\includegraphics[width=.31\textwidth]{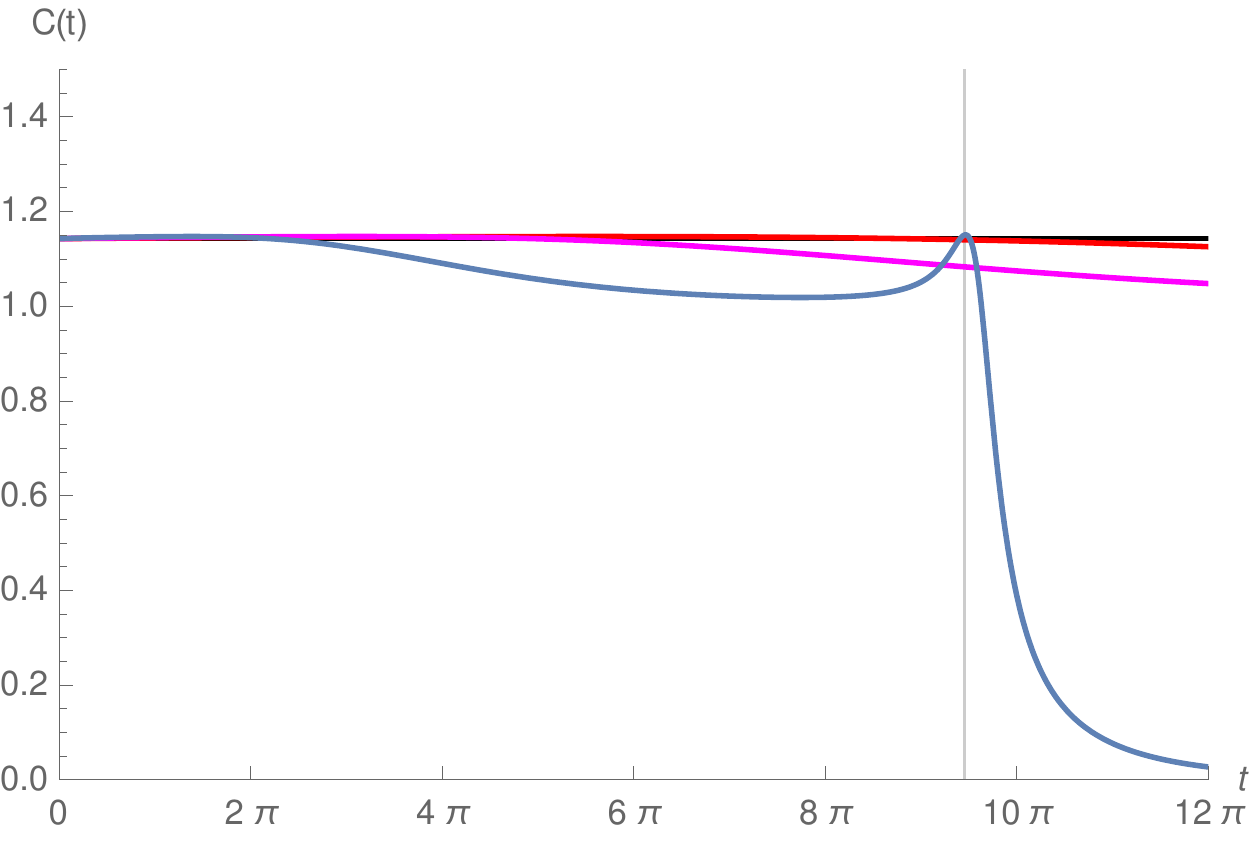}} 	\hspace{.2cm}
	\subfloat[0.996]{\includegraphics[width=.31\textwidth]{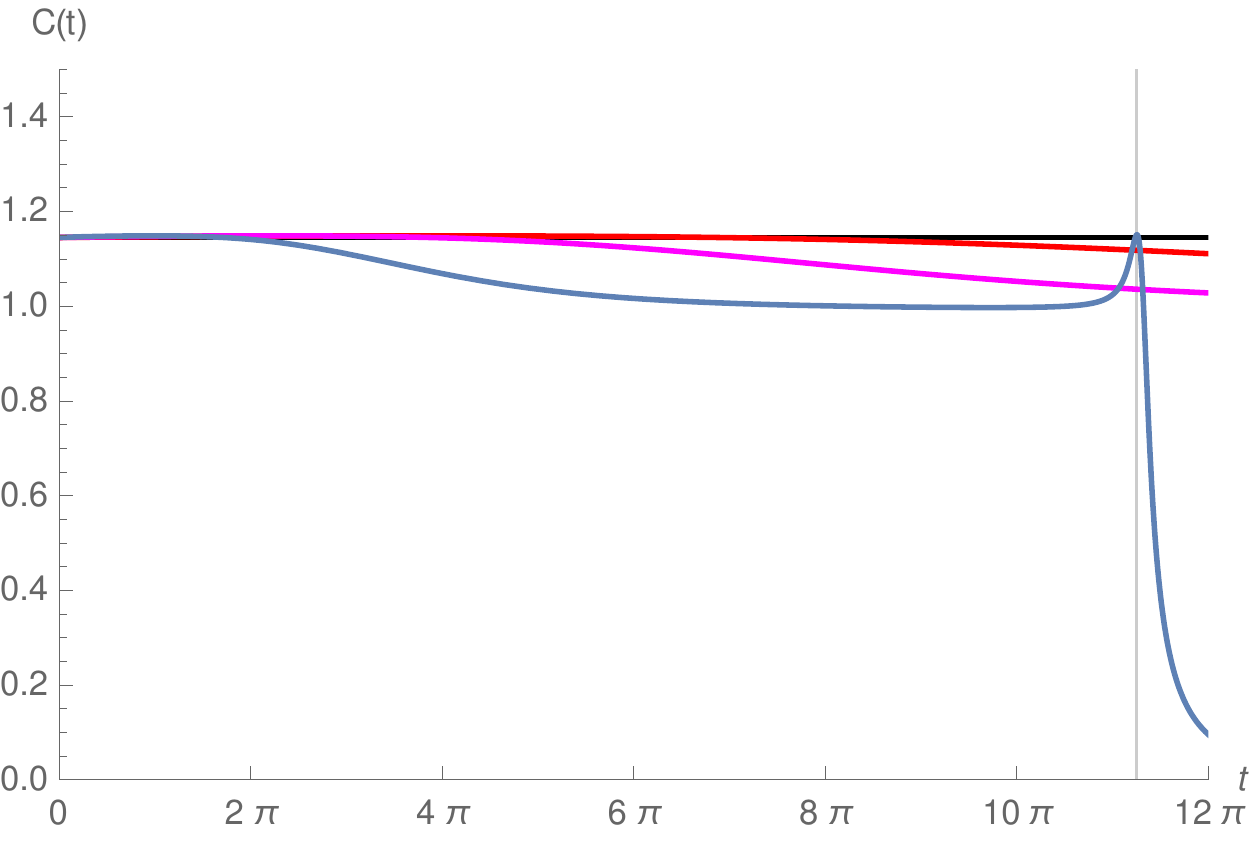}}
	\hspace{0mm} 
	\subfloat[0.997]{\includegraphics[width=.31\textwidth]{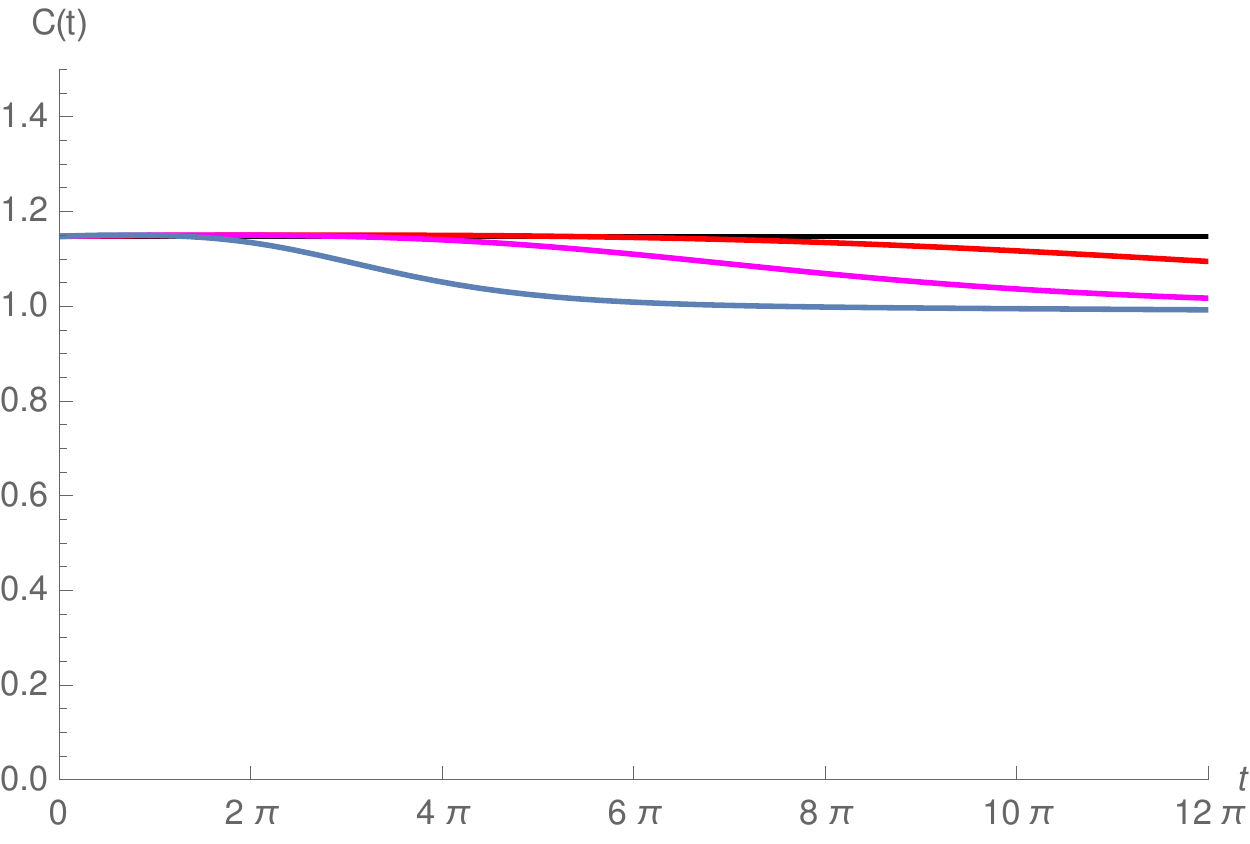}} \hspace{.2cm} 
	\subfloat[1, MES]{\includegraphics[width=.31\textwidth]{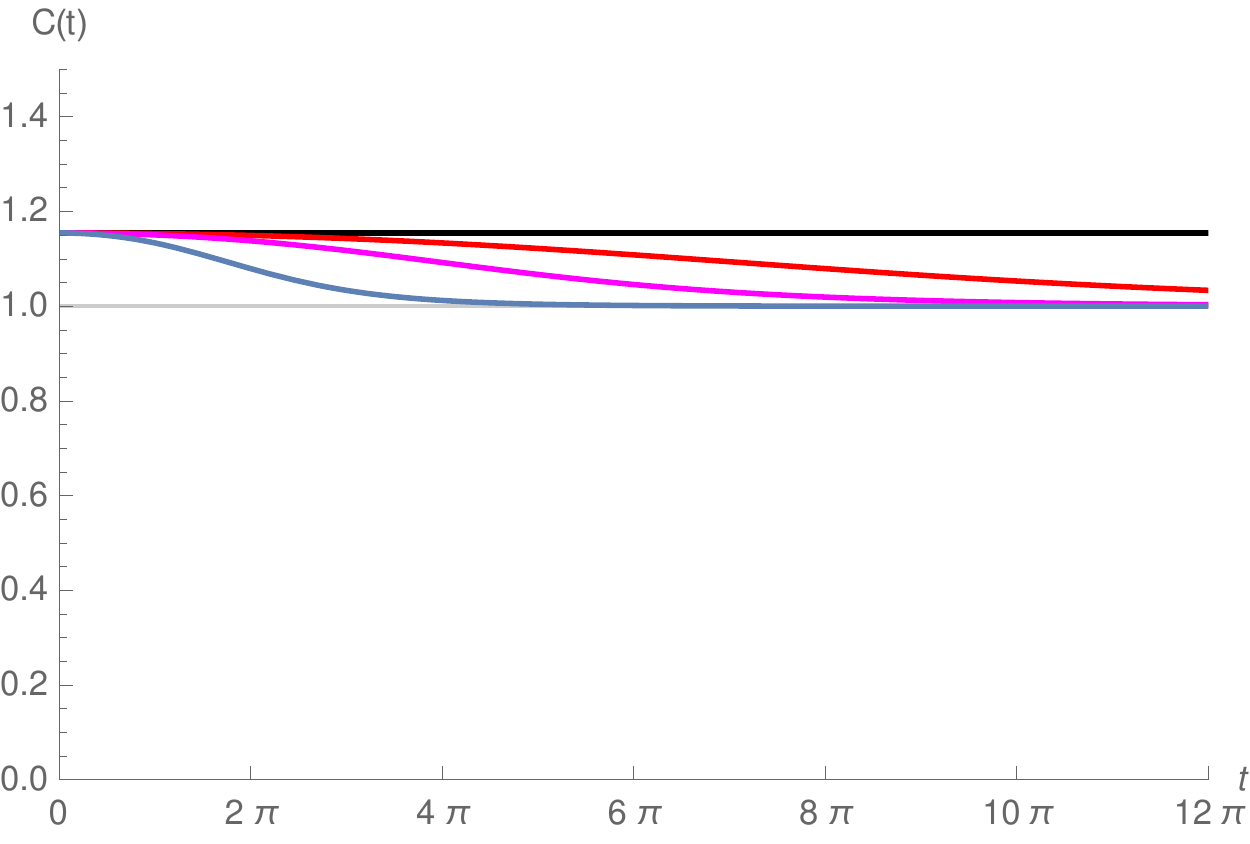}} 
\caption{Detail of the effective concurrence $C(t)$ as the initial state approaches maximal entanglement. The plots are labeled with the value of $a_0$ considered. Different couplings $(\zeta_1, \zeta_2)$ are shown as: $(0 , 0)$ in black, $(0.15 , 0.15)$ in red, $(0.2 , 0.2)$ in magenta, and $(0.3 , 0.3)$ in blue. 
 For an initial MES state, and nontrivial couplings, $C(t)$ tends to $1$ (e). }   
\label{approach-C}  
\end{figure}

So let us study in detail what happens when we continue increasing the initial concurrence.
In Fig.  \ref{approach-C}, we present the numerical results as $a_0$ changes from $0.994$ up to $1$, where the initial state becomes a MES state and $C_0$ attains its maximum value $C_m = 2/\sqrt{3} \approx 1.155$.  When $a_0$ takes the values $0.994$  (a),  
$0.995$  (b), or $0.996$  (c), the formerly undistinguishable curves start diferentiating and the curve for $(0.3,0.3)$ shows a ``kink'' localized in time (a similar behavior is expected for the other couplings, but at larger times). The kink represents an initial decay of $C(t)$ to a nonzero value, then it grows up to $C(0)$, followed by a decay to zero at later times. As the initial concurrence is increased,  the lapse of time where the kink is localized moves to larger times. For $a_0=0.997$ (d) it occurs at $t> 12 \pi$ and for the MES state ($a_0=1$)  at $t \to \infty$, leaving for finite times the decay to a nonzero value (e). This is possible as in this case we have,
 \begin{equation}
  \sigma(t) = \frac{1}{3}\left(  
\begin{array}{ccc}  
1  &  -\,e^{-\frac{t}{2} (\zeta_1 - \sqrt{3}\zeta_2)^2}  &   -\, e^{-\frac{t}{2} (\zeta_1 + \sqrt{3}\zeta_2)^2} \\   
-\, e^{-\frac{t}{2} (\zeta_1 - \sqrt{3}\zeta_2)^2}  & 1  & e^{- 2 t\, \zeta_1^2}  \\
-\, e^{-\frac{t}{2} (\zeta_1 + \sqrt{3}\zeta_2)^2} & e^{- 2 t\, \zeta_1^2}  & 1 \end{array} \right)   \;. 
\label{si-mes}
 \end{equation}
Then, the asymptotic form is $\tilde{\rho}_A= \sigma_A = \frac{1}{3}\, I$, and the spectrum is completely degenerate. Therefore, we cannot use the previous arguments to conclude that the large $t$-behavior of $u(t)$ has zero concurrence.

\begin{figure}[h]
\centering 
	\subfloat[$u_0$]{\includegraphics[width=.31\textwidth]{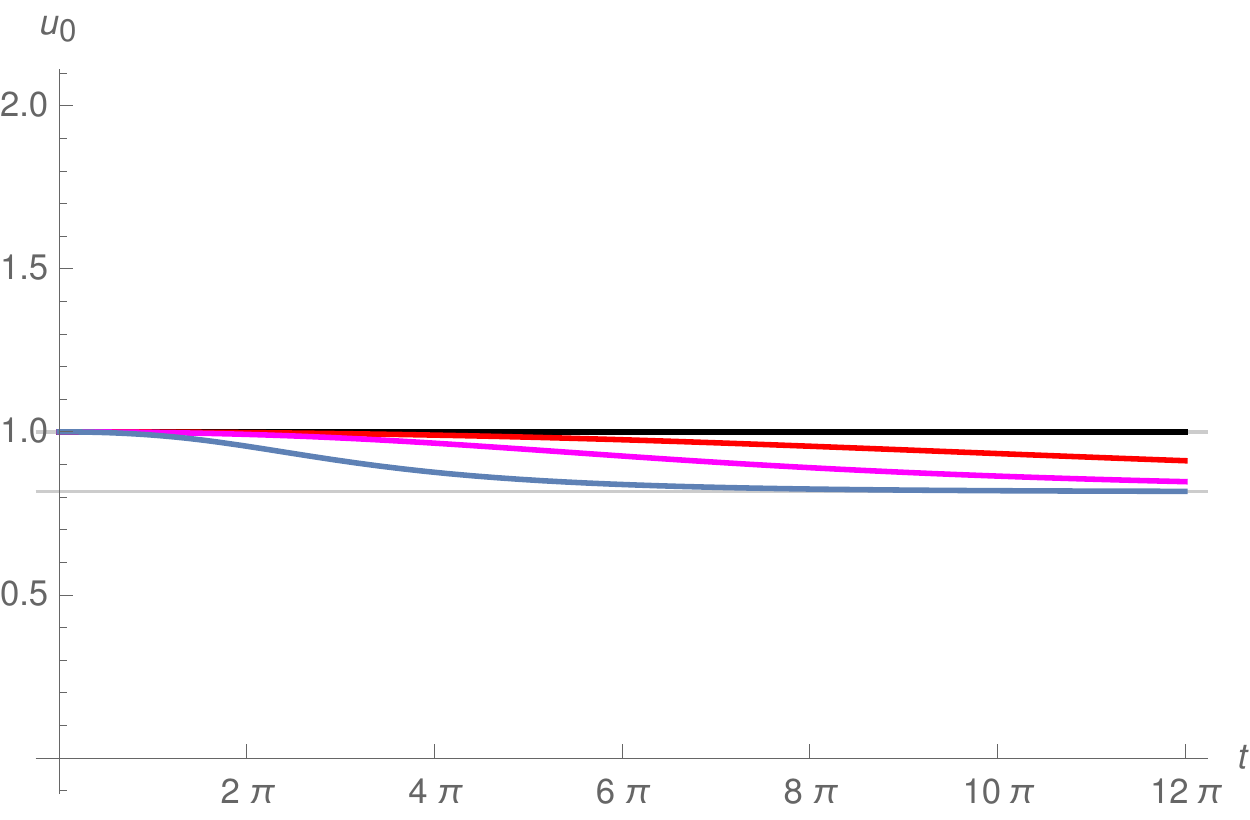}} \hspace{.2cm} 
	\subfloat[$u_1$ ]{\includegraphics[width=.31\textwidth]{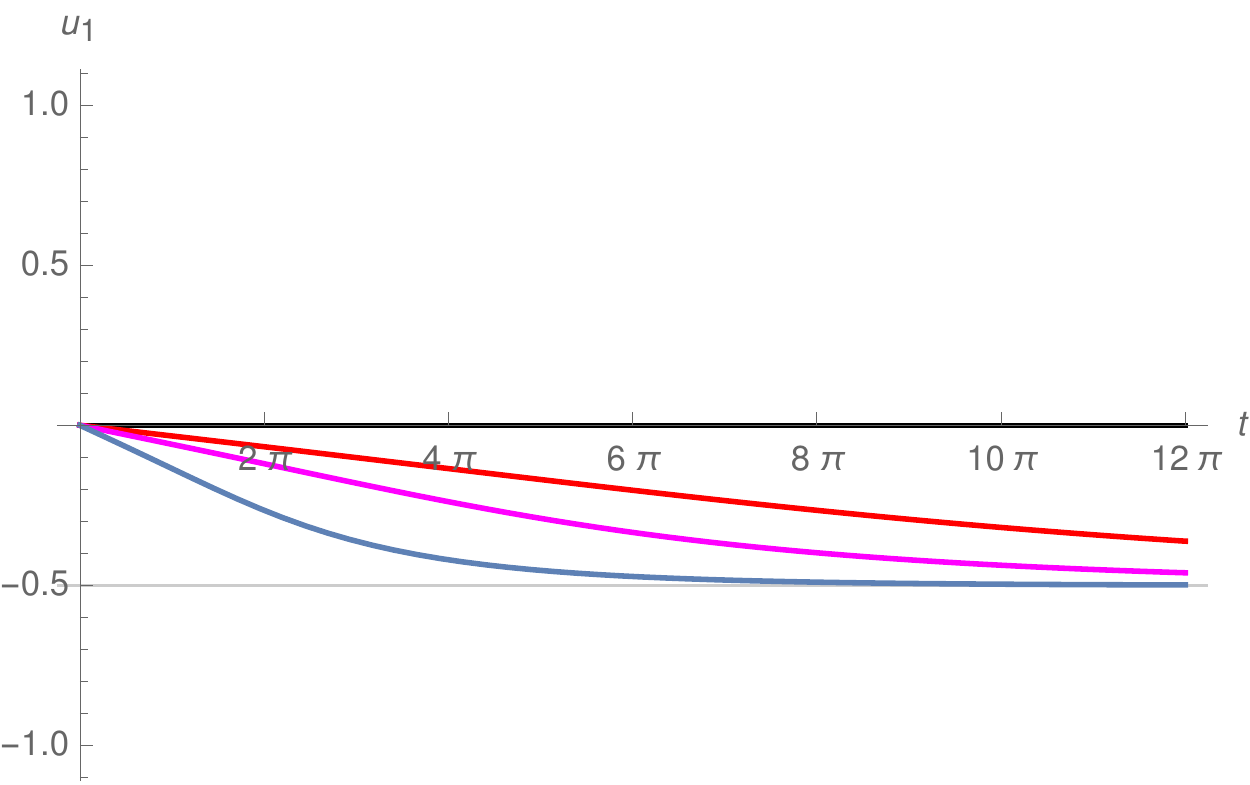}} \hspace{.2cm}
	\subfloat[$u_2$]{\includegraphics[width=.31\textwidth]{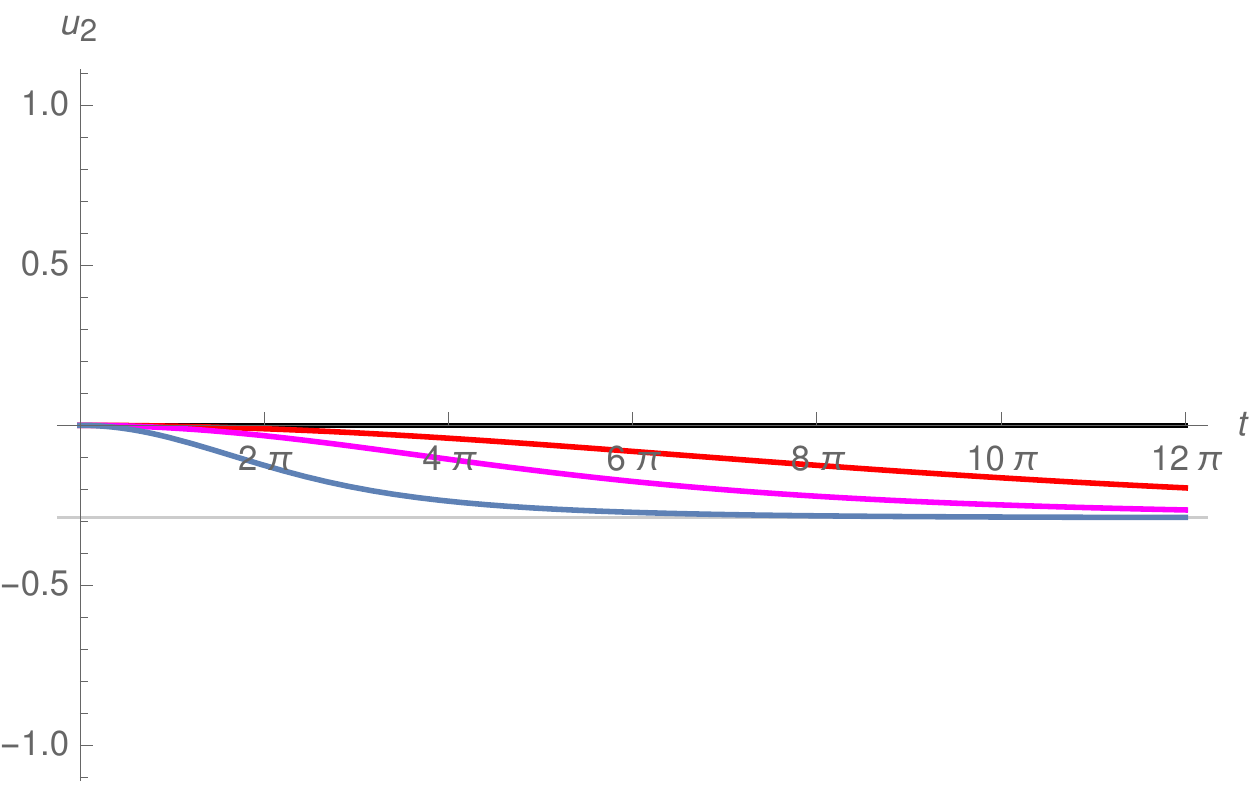}}  
\caption{Time dependence of $u_0$, $u_1$ and $u_2$ for an initial MES: $(0 , 0)$ in black, $(0.15 , 0.15)$ in red, $(0.2 , 0.2)$ in magenta, $(0.3 , 0.3)$ in blue. 
For the nontrivial couplings, they tend to $\sqrt{\frac{2}{3}}$ in (a), $-\frac{1}{2}$ in (b), and $-\frac{1}{\sqrt{12}}$ in (c).}   
\label{asy-coef} 
\end{figure}

Now, it is important to note that the finite asymptotic value when the system starts from a  MES state happens to be $1$ (\ref{approach-C}-e). This points to the asymptotic 
effective state as a maximally entangled two-qubit object. Indeed, Fig.  \ref{asy-coef} shows the numerical result for the time-dependence of the coefficients  $u_0$ (a), $u_1$ (b), and $u_2$ (c) of the effective state, for a pure initial MES state. The plots display the asymptotic values, given by $\sqrt{2/3}$, $-1/2$, and $-1/\sqrt{12}$, respectively.
That is, the  environment drives the initial state $|0\rangle$ into an effective state 
$|\Psi'(t)\rangle $, with eigenvalue $\varepsilon(t)$, and asymptotic behavior,
\[
|0 \rangle  \to |\Psi'(t_{\rm A})\rangle \sim \sqrt{\frac{2}{3}}\,  |0\rangle -\frac{1}{2} \,  |1\rangle -\frac{1}{\sqrt{12}} \, |2\rangle \;.
\] 
In matrix representation (cf Eq. (\ref{pprima})) this corresponds to, 

\begin{equation} 
\Psi'(0)=\frac{1}{\sqrt{3}}\left(  
\begin{array}{ccc}    
1  & 0 &  0 \\   
0 & 1  & 0  \\
0 & 0 & 1 \end{array} \right)  \to \Psi'(t_{\rm A}) \sim \frac{1}{\sqrt{2}}\left(  
\begin{array}{ccc}    
0  & 0 &  0 \\   
0 & 1  & 0  \\
0 & 0 & 1 \end{array} \right)   \;,
\end{equation}
while in the $|ij\rangle$ basis, we have,

\[
\frac{1}{\sqrt{3}}\, (  |11\rangle + |22\rangle + |33\rangle ) \to  \frac{1}{\sqrt{2}}\, (  |22\rangle + |33\rangle )  \;.
\]

\begin{figure}[h]
\includegraphics[scale=0.6]{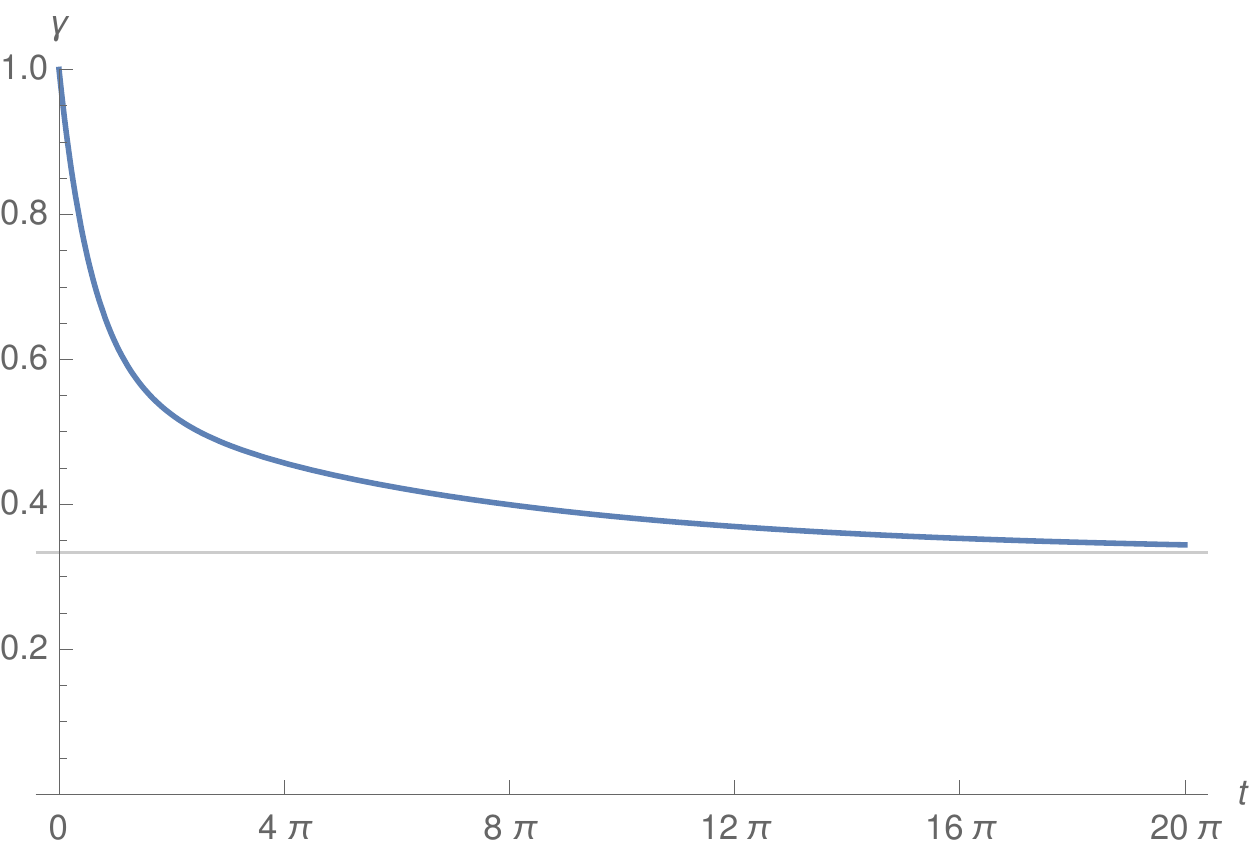} 
\caption{\label{} Purity for an initial MES state. The asymptote is at $\gamma =1$, as expected. } 
\label{Purity}
\end{figure}

The other instantaneous eigenvectors $| \Psi'_1(t) \rangle$, $| \Psi'_2(t) \rangle$ (with eigenvalues $\varepsilon_1(t)$, $\varepsilon_2(t)$) undergo the following initial to asymptotic transitions, observed numerically,  

\[
\frac{1}{\sqrt{2}}\, \left(|1\rangle + |2\rangle \right) \to |\Psi'_1(t_{\rm A}) \rangle \sim  \frac{1}{\sqrt{3}} \, |0\rangle +  \frac{1}{\sqrt{2}} \, |1\rangle + \frac{1}{\sqrt{6}} \, |2\rangle \]
\[
\frac{1}{\sqrt{2}}\, (|1\rangle - |2\rangle ) \to  |\Psi'_2(t_{\rm A}) \rangle \sim \frac{1}{2} \, |1\rangle - \frac{\sqrt{3}}{2} \, |2\rangle  \;,
\]
 which in matrix  form correspond to, 
\begin{equation} 
\frac{1}{2\sqrt{3}}
\left(  
\begin{array}{ccc}    
 1+ \sqrt{3}  & 0 &  0 \\   
0 &   1 -\sqrt{3}  & 0  \\
0 & 0 & - 2 \end{array} \right) \to \left(  
\begin{array}{ccc}    
 1  & 0 &  0 \\   
0 &  0  & 0  \\
0 & 0 & 0  \end{array} \right) 
\end{equation}
\begin{equation}
-\frac{1}{2\sqrt{3}}\left(  
\begin{array}{ccc}    
 1-\sqrt{3}  & 0 &  0 \\   
0 &   1 +\sqrt{3}  & 0  \\
0 & 0 &  -2 \end{array} \right) 
\to
\frac{1}{\sqrt{2}}\left(  
\begin{array}{ccc}    
0  & 0 &  0 \\   
0 & -1  & 0  \\
0 & 0 & 1 \end{array} \right) 
  \;,
\end{equation}
and in the $|ij\rangle $ basis,
\[
\frac{1}{2\sqrt{3}}\, ( ( 1+ \sqrt{3})\, |11\rangle +  (1 - \sqrt{3})\,|22\rangle -2\, |33\rangle ) \to     |11\rangle \;,
\] 
\[
-\frac{1}{2\sqrt{3}}\, ( ( 1 - \sqrt{3})\, |11\rangle +  (1 + \sqrt{3})\,|22\rangle - 2\, |33\rangle ) \to  \frac{1}{\sqrt{2}}\,  (  |33\rangle - |22\rangle )  \;,
\] 
respectively. Let us consider a global property of the two-qutrit system, namely, the evolution of purity $\gamma = {\rm  Tr}\,[\tilde{\rho}^2]\,$ restricted to the diagonal sector. 
It can be expressed in terms of the eigenvalues of the density matrix, in our case,

 \[
 \gamma = [\varepsilon(t)]^2 + [\varepsilon_1(t)]^2 + [\varepsilon_2(t)]^2   \;.
 \] 
In Fig.  \ref{Purity}, we display it  when starting from a MES state.
It evolves from $1$, for the initial pure state, to the minimum value $1/3$ allowed for this restricted sector. Note that the global minimal purity is $1/9\,$, corresponding to a two-qutrit completely mixed state. 

\subsection{Underlying eigenvalue properties }

\begin{figure}[h]
\centering  
		\subfloat[0.90]{\includegraphics[width=.31\textwidth]{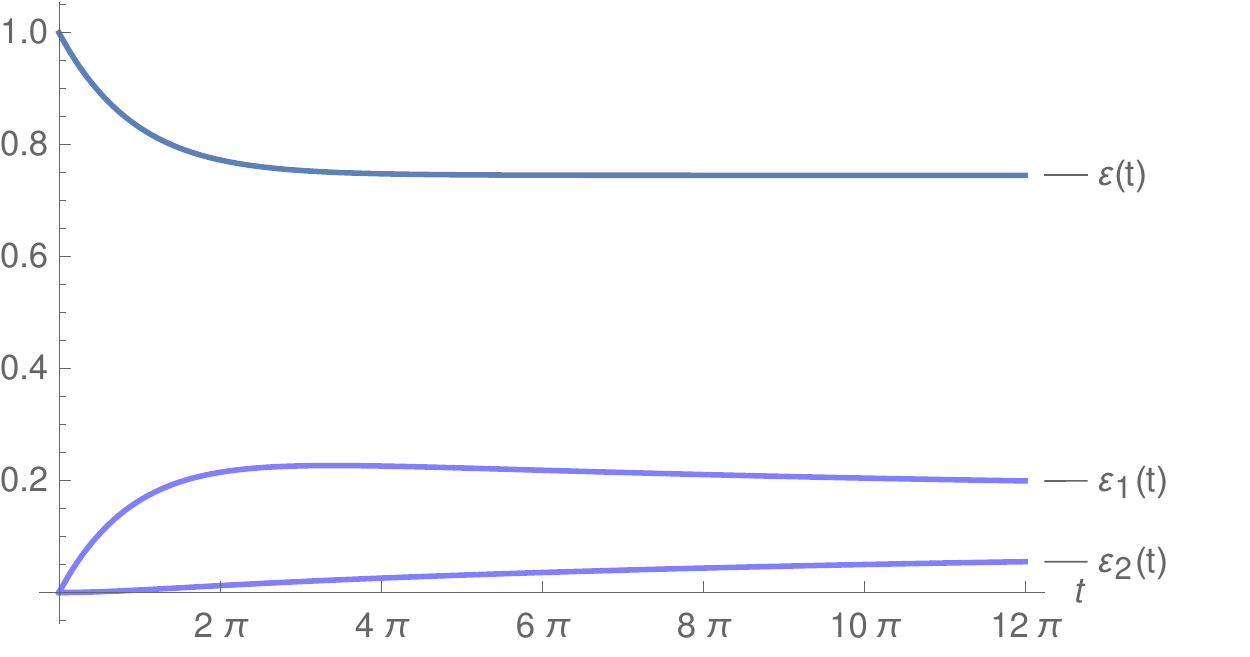}} 
		\hspace{.2cm} 
 	\subfloat[0.99]{\includegraphics[width=.31\textwidth]{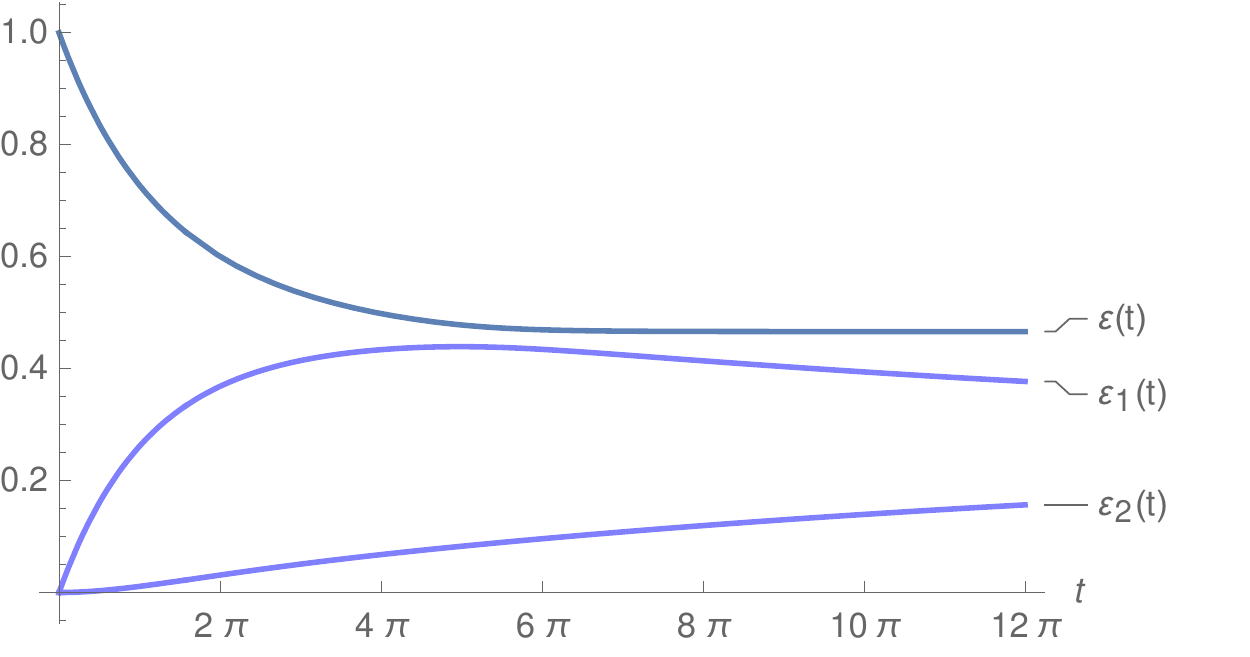}} 
 	  \hspace{.2cm} 
 		\subfloat[0.993]{\includegraphics[width=.31\textwidth]{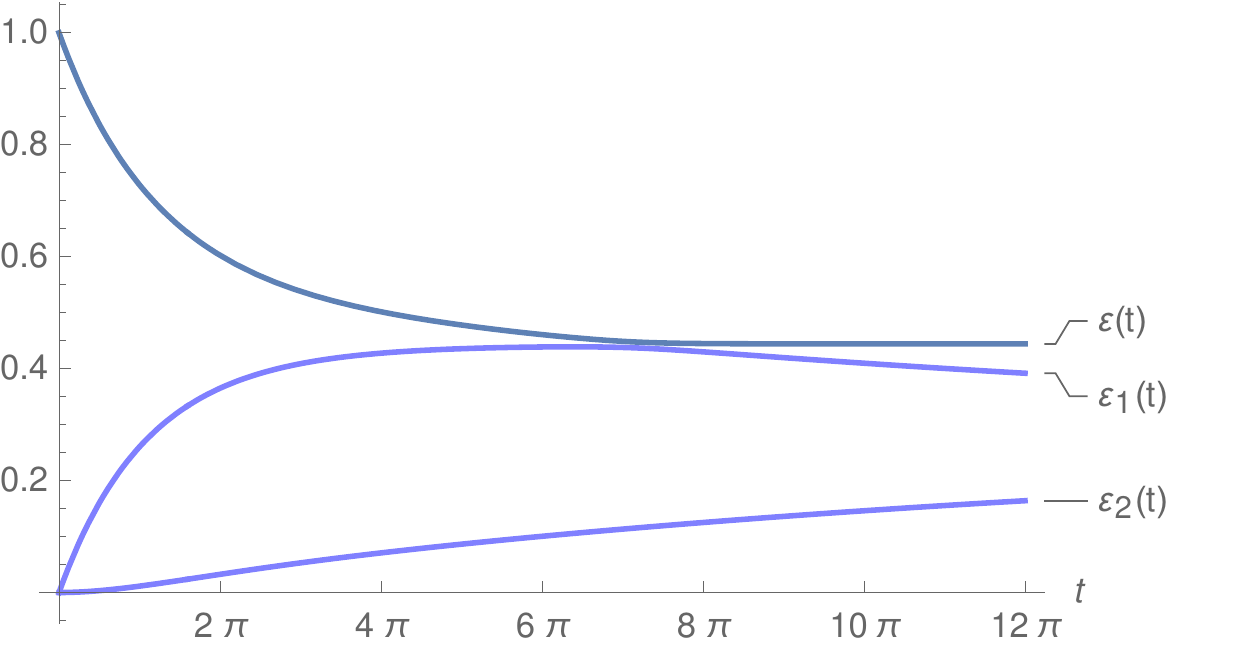}}
\caption{ The eigenvalues of the density matrix.  The plots are labeled with the value of $a_0$ considered.  The couplings are fixed at $(\zeta_1, \zeta_2)=(0.3 , 0.3)$. }  
\label{Eig-2}   
\end{figure}

\begin{figure}
\centering  
	\subfloat[0.994]{\includegraphics[width=.31\textwidth]{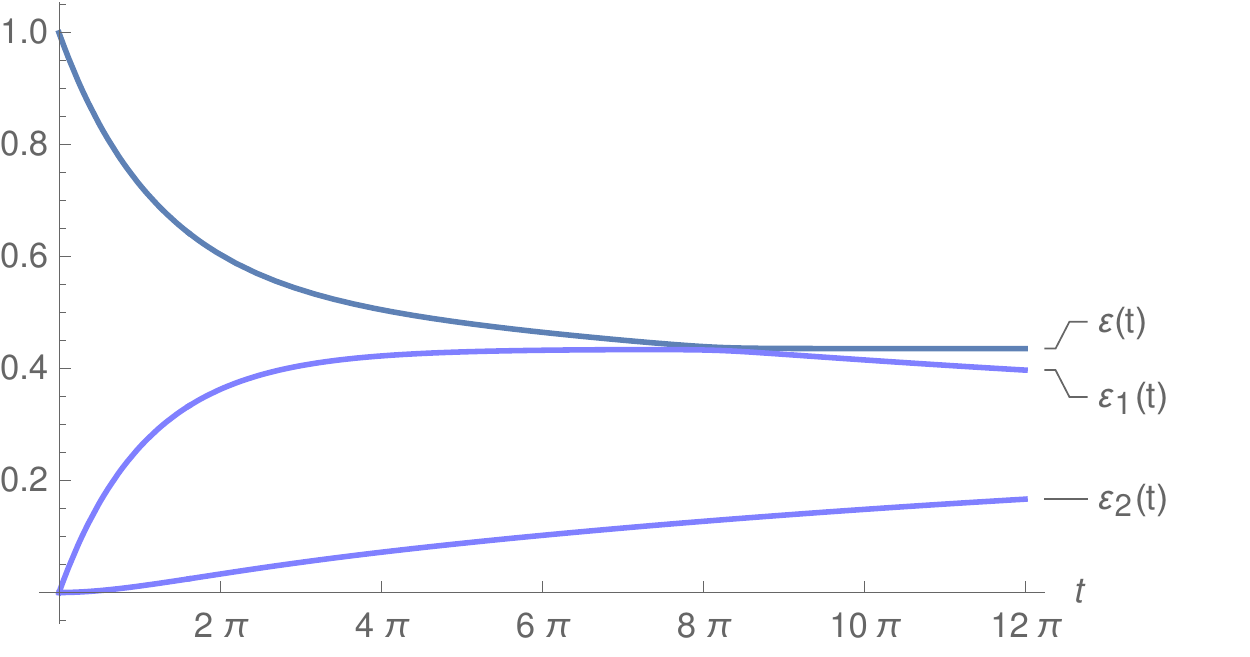}} 
	 \hspace{.2cm} 
	\subfloat[0.995]{\includegraphics[width=.31\textwidth]{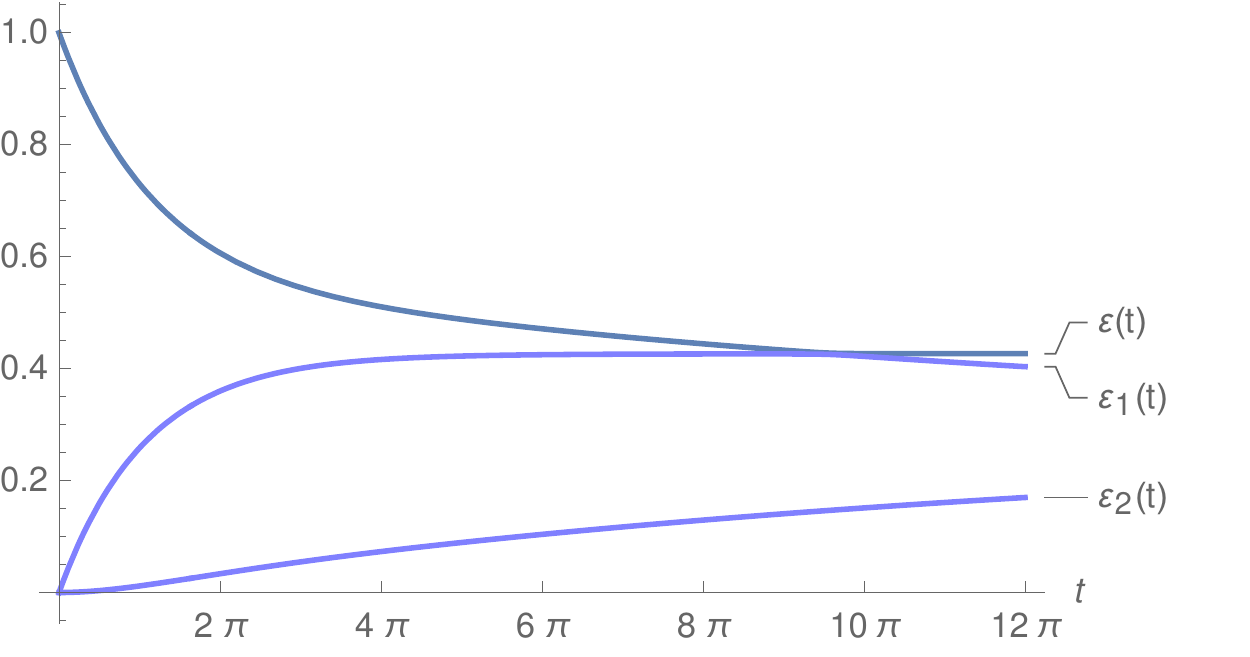}} 
	\hspace{.2cm} 	
	\subfloat[0.996]{\includegraphics[width=.31\textwidth]{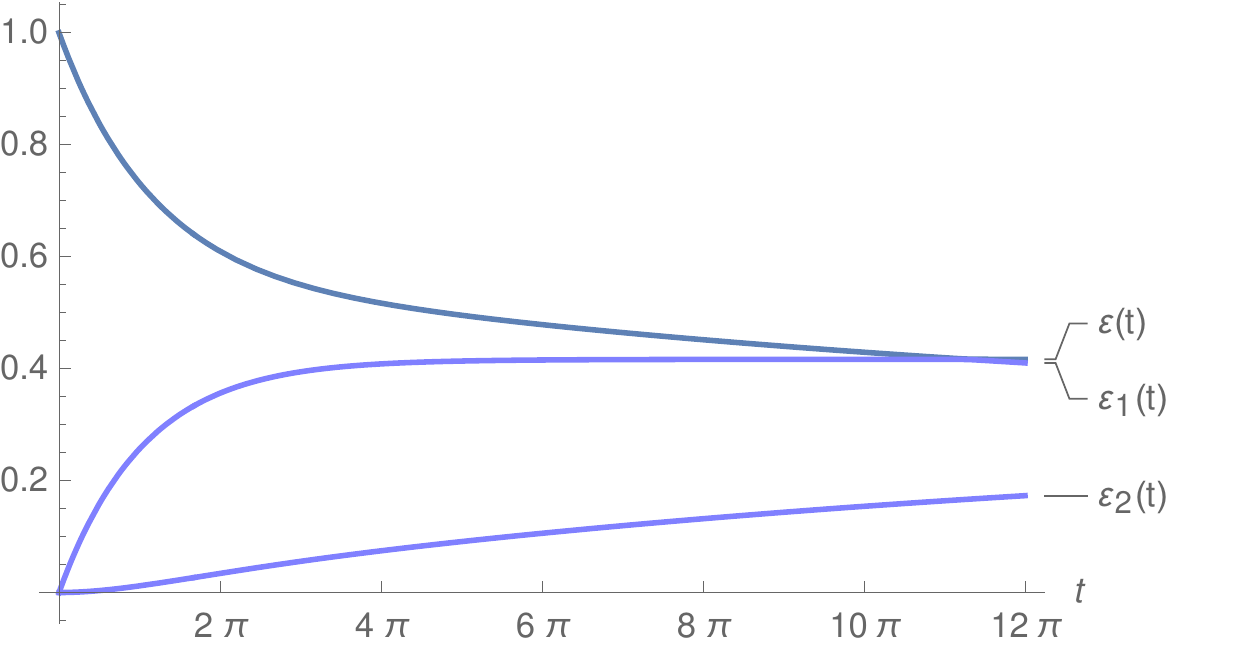}}
	\hspace{0mm} 
	\subfloat[0.997]{\includegraphics[width=.31\textwidth]{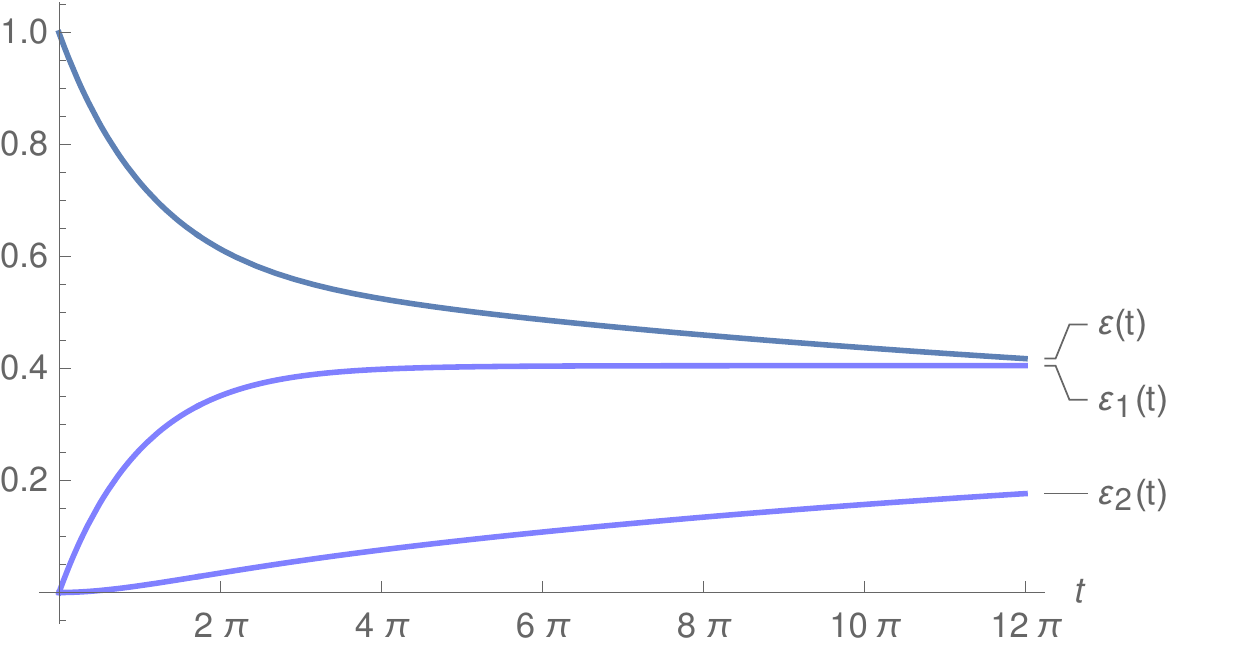}} 
	\hspace{.2cm} 
	\subfloat[MES]{\includegraphics[width=.31\textwidth]{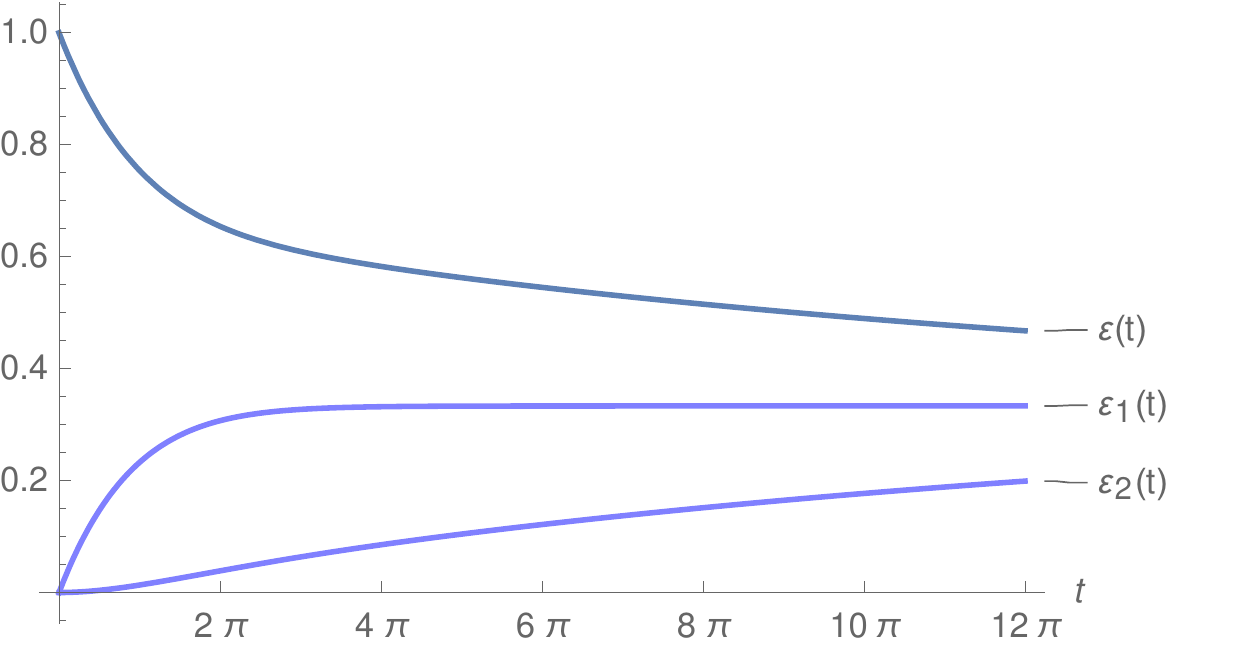}} 
\caption{ Detail of the eigenvalues as the initial state approaches maximal entanglement. The plots are labeled with the value of $a_0$ considered. The couplings are fixed at $(\zeta_1, \zeta_2)=(0.3 , 0.3)$. }   
\label{Eig-6}  
\end{figure}

\begin{figure}
\centering 
	\subfloat[0.994]{\includegraphics[width=.31\textwidth]{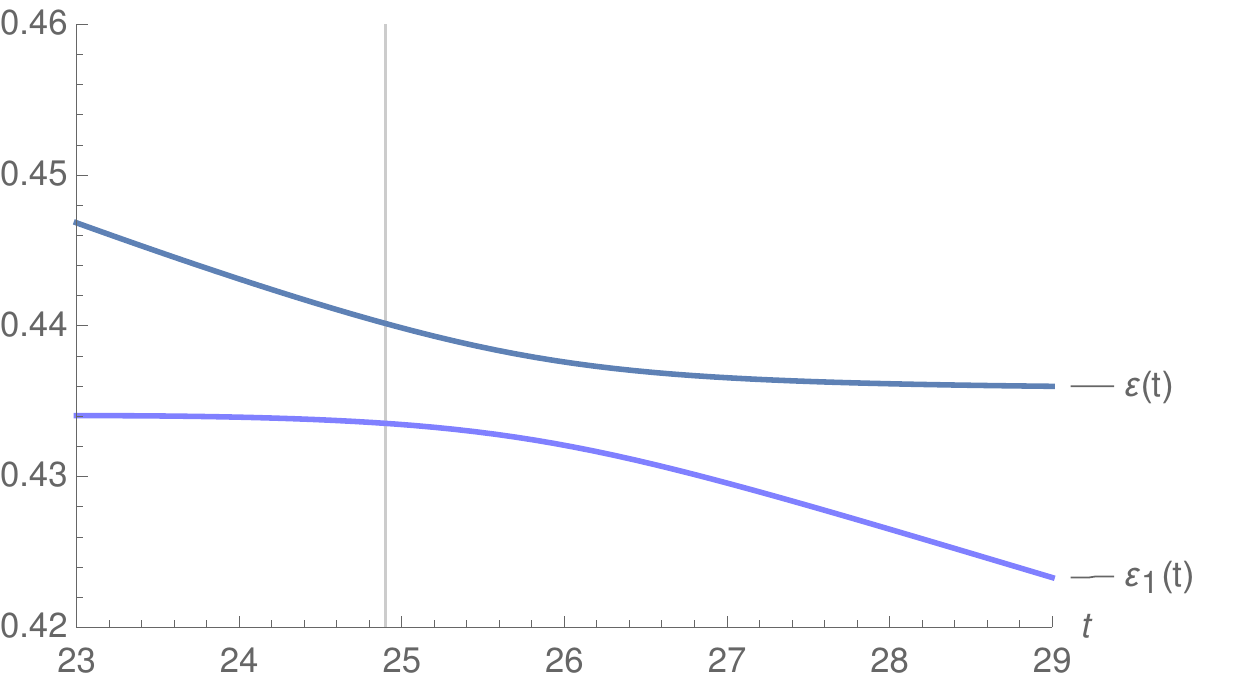}}    
	\hspace{.2cm} 
	\subfloat[0.995]{\includegraphics[width=.31\textwidth]{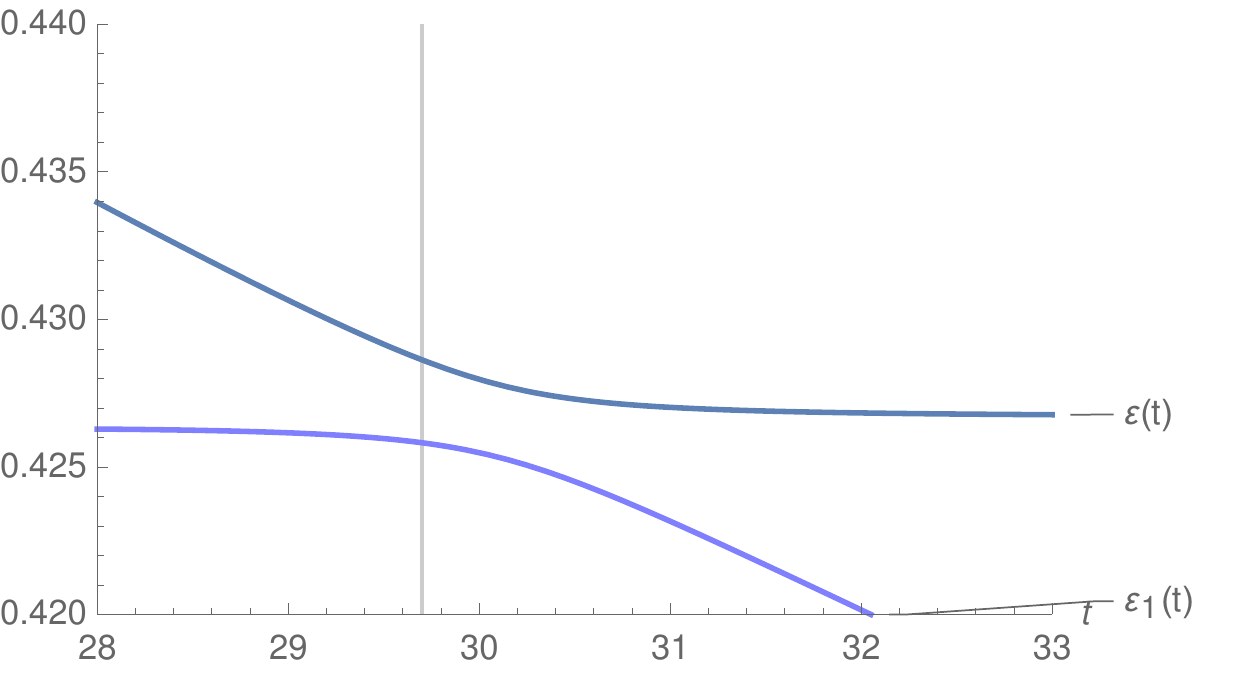}} 
	\hspace{.2cm} 
	\subfloat[0.996]{\includegraphics[width=.31\textwidth]{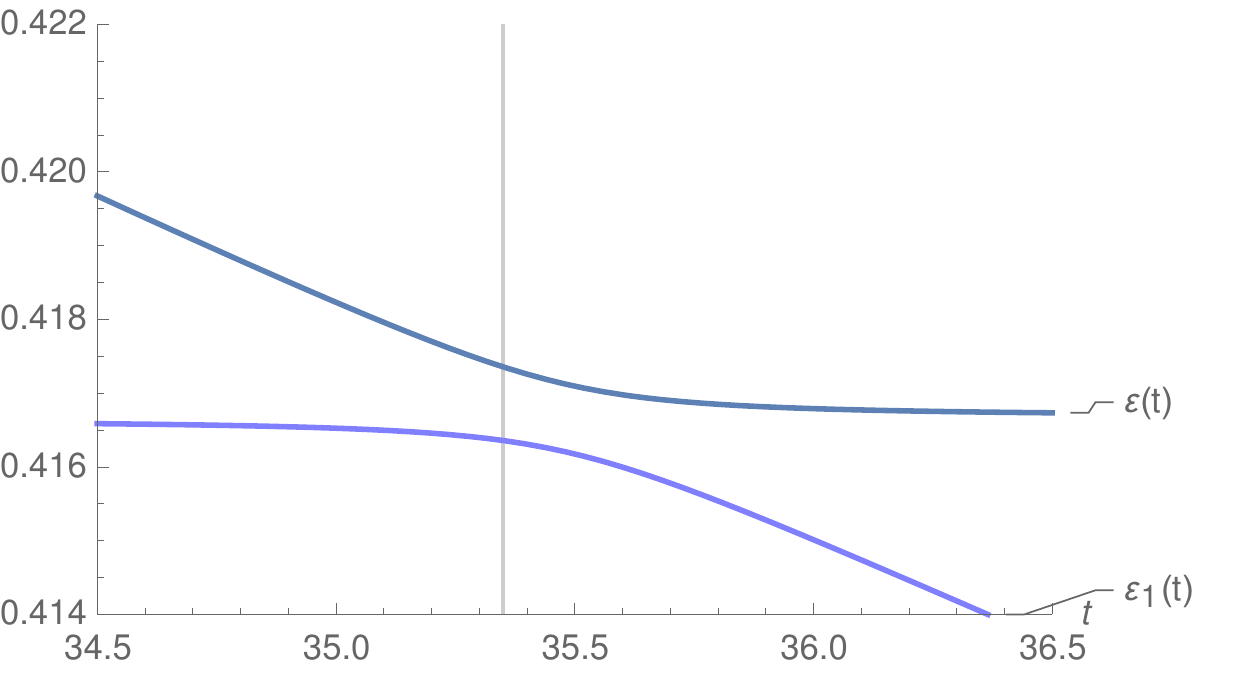}}
\caption{A zoom for the plots in Fig. \ref{Eig-6}, in the regions where the eigenvalues $\varepsilon(t)$ and $\varepsilon_1(t)$ approach each other.    The plots are labeled with the value of $a_0$ considered. The couplings are fixed at $(\zeta_1, \zeta_2)=(0.3 , 0.3)$.}   
\label{zoom}  
\end{figure}

 In this section, we would like to shed some light on the behavior of the effective concurrence $C(t)$ as a function of the initial sate (Fig.  \ref{approach-C}).
As this concept  applies to the effective state, which is a particular eigenstate, let us turn our attention to the evolution of the individual eigenvalues. 
In Figs. \ref{Eig-2} and \ref{Eig-6},  we show the time-dependence of the eigenvalues of the density matrix. All of them were generated for $(\zeta_1, \zeta_2)=(0.3 , 0.3)$. 
 The plots for different
$a_0$ values are in one-to-one correspondence with the blue lines for the effective concurrence in   Figs. \ref{conj1} and \ref{approach-C}. We note that the change from a normal decay (Fig. \ref{conj1}a) to the persistent  behaviors (Figs. \ref{conj1}b, \ref{conj1}c), is accompanied by a change from well-separated eigenvalue evolutions  
(\ref{Eig-2}a) to a situation where two of them get closer on some time interval (Figs. \ref{Eig-2}b,  \ref{Eig-2}c). In addition, the instant when a kink gets a maximum  (Figs.\ref{approach-C}a, \ref{approach-C}b, \ref{approach-C}c) is correlated with the 
instant when the eigenvalues become nearly degenerate ((Figs. \ref{Eig-6}a, \ref{Eig-6}b, \ref{Eig-6}c)). This can be clearly seen in 
Fig. \ref{zoom}. There, we considered the vertical line for each peak in Fig. \ref{approach-C}, and verified that it passes on the corresponding zoomed region where the eigenvalues attain their maximum approximation.  
During the close approach between the eigenvalues $\epsilon(t)$ and
$\epsilon_{1}(t)\,$, the associated eigenvectors are close to $(|22\rangle + | 33\rangle ) /\sqrt{2}$ and
$| 11\rangle $, spanning a quasi-degenerate subspace. This allows the
transient formation of a state that is close to the maximally entangled eigenstate 
$ (\left| 11\rangle + | 22\rangle\ + | 33\rangle\right)/\sqrt{3}$, giving rise
to the concurrence peak. As the eigenvalues diverge from each other,
this superposition is broken and the individual components continue
the decoherence process, following a rapid decay. 
 
\section{Geometric phase for an initial qutrit MES state}
\label{GPmes}

If the initial state is separable, say $|11 \rangle \langle 11|$, we have $(a_0, a_1, a_2) =(\frac{1}{\sqrt{3}},\frac{1}{\sqrt{2}},\frac{1}{\sqrt{6}} )$, and $(b_0,b_1, b_2) = (0,0,1)$. In this case, $\sigma(0)$ is diagonal so that, $\partial_t \sigma|_0 =  - [D, [D,\sigma ]]|_0 =0$, and $\sigma(t) \equiv \sigma(0)$. Then, the system remains pure, in a separable state, and the geometric phase changes linearly. 

In general, for times larger than the typical asymptotic time ($t > t_{\rm A}$), the geometric phase becomes,

\begin{eqnarray}
\phi_g (t) &=& \arg \, {\rm Tr}  \left[ (\Psi'(0))^\dagger \,  U_{\rm S}(t)\, \Psi'(t_{\rm A}) \right]  
\nonumber \\
&+&  
\int_0^{t_{\rm A}} ds \,\, {\rm Tr}  \left[ (\Psi'(s))^\dagger  H_{\rm S} \,\Psi'(s) \right] +
(t-t_{\rm A}) \, {\rm Tr}  \left[ ({\Psi'}(t_{\rm A})^\dagger  H_{\rm S} \,\Psi'(t_{\rm A}) \right] \;. 
\end{eqnarray}    

\begin{figure}[h]
\centering 
	\subfloat[Operating with $\vec{w}_1$]{\includegraphics[width=.49\textwidth]{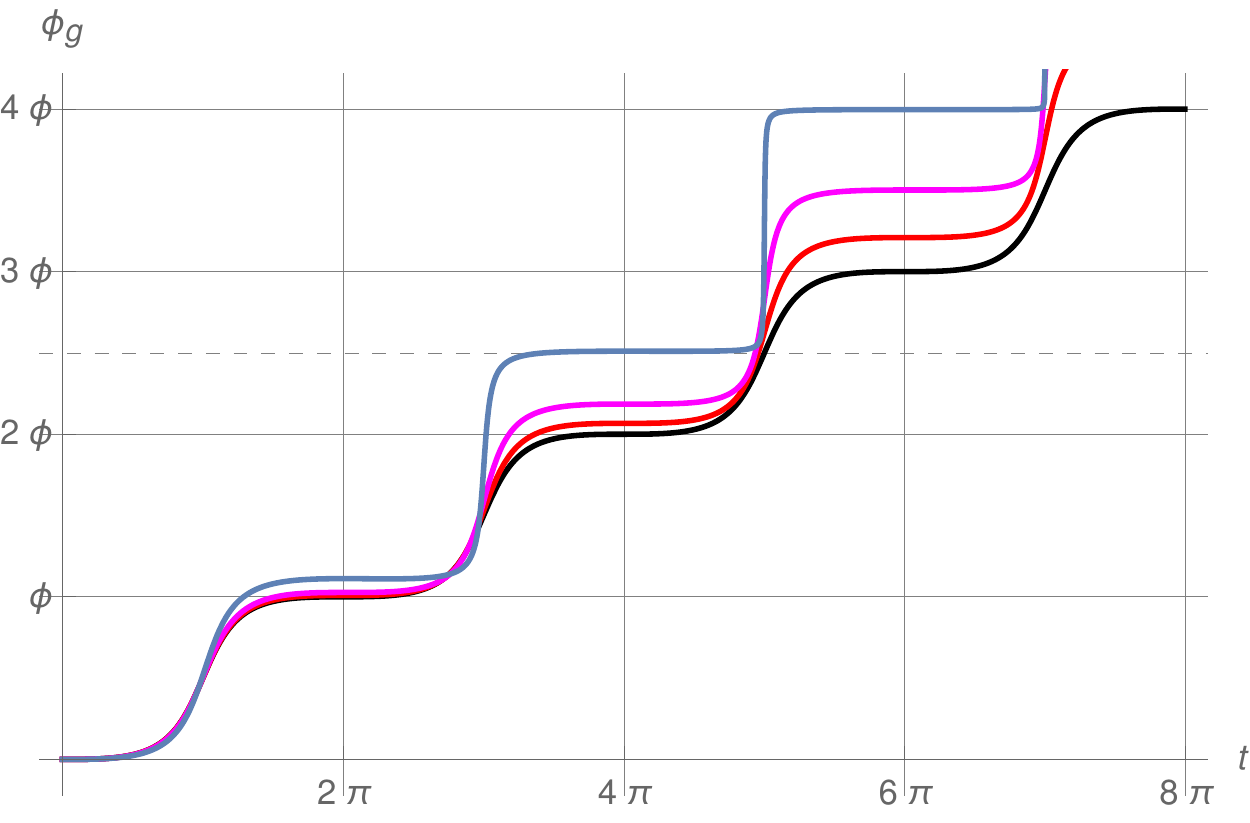}} \hfill
	\hspace{0mm}
	\subfloat[Operating with $\vec{w}_3$]{\includegraphics[width=.49\textwidth]{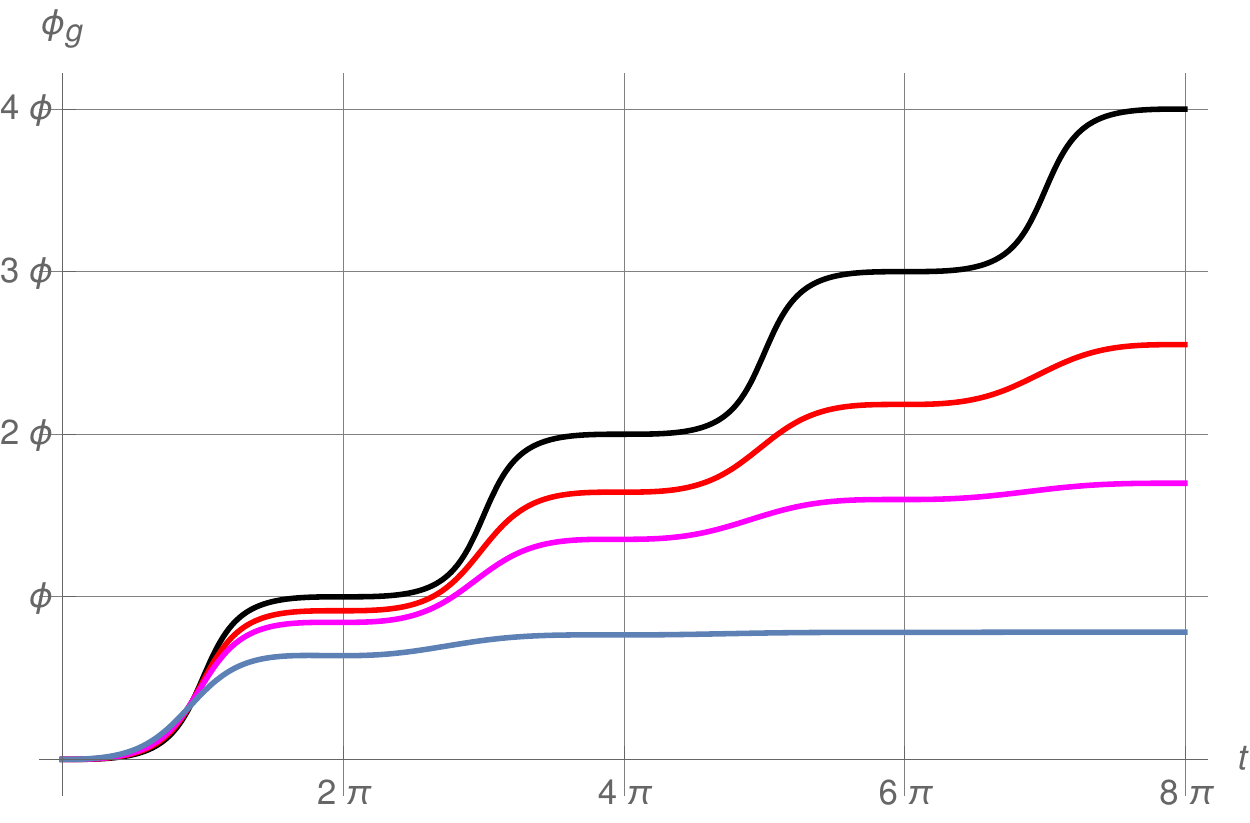}} \hfill
\caption{The effect of the environment on the geometric phase, starting from a MES state operated on the first qubit. The couplings $(\zeta_1, \zeta_2)$ are: (0,0) (in black), (0.15,0,15) (in red), (0.20,0,20) (in magenta), and (0.30,0,30) (in blue) (this ordering corresponds to bottom to top (a) and top to bottom (b)), $\phi=\frac{2\pi}{3}$.}     
\label{geo}  
\end{figure}

Now, let us take a closer look to the geometric phase $\phi_g(t)$ for a MES state. In this case, we have seen that the effective state, needed to compute this phase, undergoes a transition from a MES two-qutrit state to an effective MES two-qubit object, driven by the environment. In Fig. \ref{geo}, we show the numerical result for 
$\phi_g(t)$ in Eq. (\ref{phasep}) using different couplings, that are in correspondence with those used to analyze the effective concurrence (Fig.  \ref{approach-C}e) and the coefficients of the effective state (Fig. \ref{asy-coef}).  Consider the system 
operated by the weight $\vec{w}_1$ (a).  For  $(\zeta_1,\zeta_2)=(0,0)$ (in black),
we have the $\phi_g$-evolution for qutrit MES state, with fractional phase jumps of $\phi = 2\pi/3$. On the other hand, for $(\zeta_1, \zeta_2)=(0.3, 0,3)$ (in blue), we clearly see an initial behavior with a jump $\approx \phi$, and a later behavior with jumps of $\approx \frac{3}{2} \phi =\pi$. Of course, for intermediate couplings, the change in the regime takes a longer time. When the system is operated with $\vec{w}_2$, a similar behavior was verified. On the other hand, when it is operated with $\vec{w}_3$ (b), the decoherence destroys the pattern.

 These properties led us to explore the parameter space, consisting in the couplings $(\zeta_1 , \zeta_ 2)$, and for a given choice, the possibility of operating the system with three different weights $\vec{w}_1$, $\vec{w}_2$ and $\vec{w_3}$. In all the cases considered, we observed: i) the concurrence of an  eigenvector of the density matrix tends to zero, while that of the other two eigenvectors tends to $1$ (the maximum concurrence for a two qubit state); ii) the concurrence of the effective state tends to $1$.
 
\begin{figure}[h]
\includegraphics[scale=0.6]{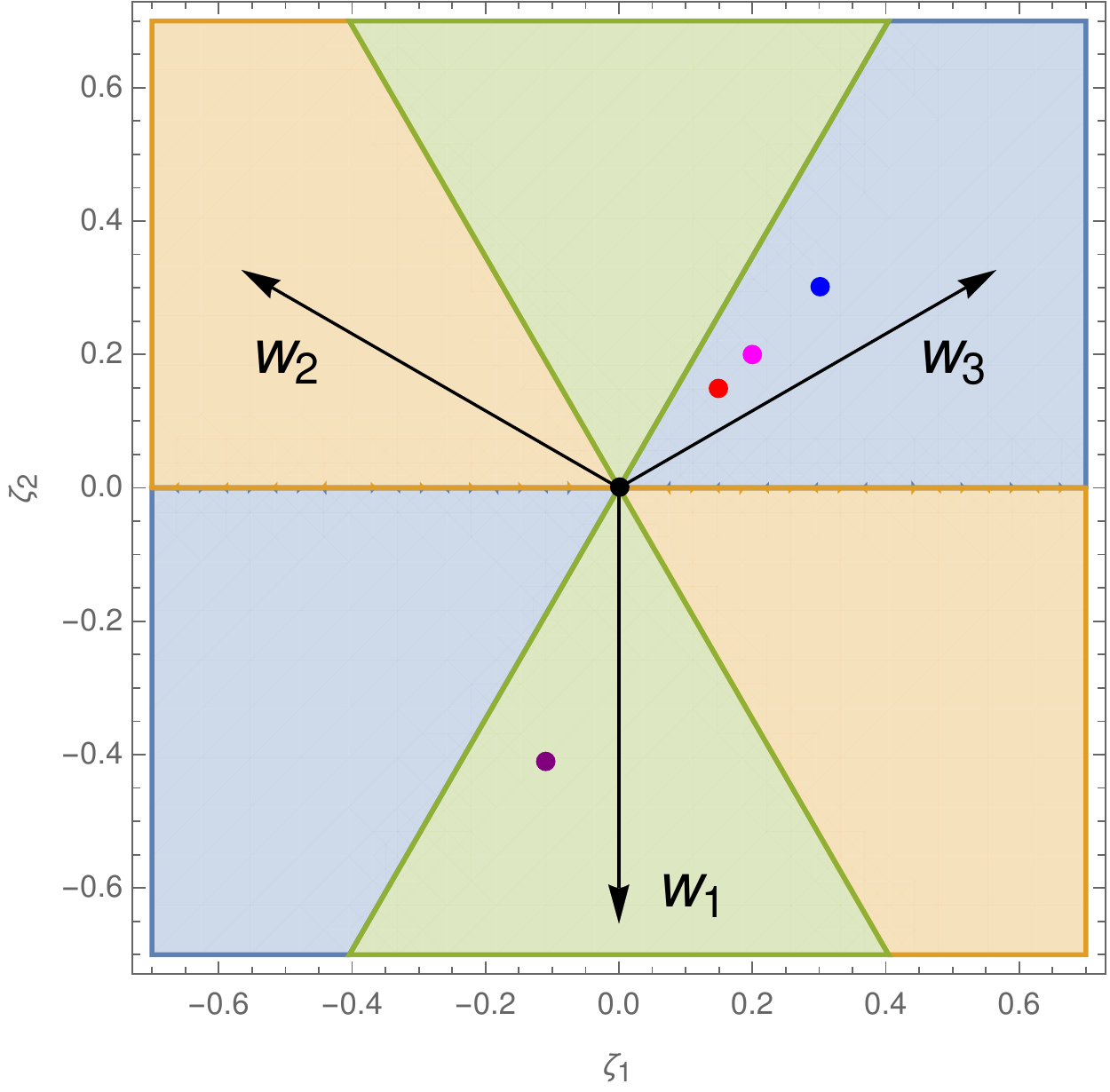}  
\caption{\label{} Regions. } 
\label{Regions}
\end{figure}

In addition, the geometric phase evolutions are either of type (a) or (b) (see Fig. \ref{geo}), depending on the relation between $(\zeta_1 ,\zeta_2 )$ and the weight used to perform the operations. In this regard, the couplings can be classified by initially denoting the exponentials in Eq. (\ref{si-mes}) as, $e^{-At/2}$, $e^{-Bt/2}$, $e^{-Ct/2}$, with $A=(\zeta_1-\sqrt{3}\, \zeta_2)^2$,   $B=(\zeta_1+\sqrt{3}\, \zeta_2)^2$, $C=(2\zeta_1)^2$. Next we define the regions where ($A< B$ and $A < C$), ($B< A$ and $B < C$), ($C< A$ and $C < B$). This regions are dominated by the exponential with the smallest decaying rate, that is, $A$, $B$, 
and $C$, respectively. In Fig. \ref{Regions}, they are shown in parameter space $(\zeta_1, \zeta_2)$, and correspond to the areas in grey, orange, and green.

In that figure, the 
indicated $\vec{w}_i$'s denote the three regions by using the contained weight. In these regions, we can still operate the system with any weight. For example, the dots in black (central), red, magenta, and blue, correspond to the couplings used in Fig. (\ref{geo}) to compute the geometric phase. As already mentioned, we operated each one of those cases with single qutrit evolutions characterized by $\vec{w}_1$,  $\vec{w}_2$ and  $\vec{w}_3$. In table I, we list the possible combinations of weights and the type of pattern observed, whether it is type (a) or (b).

\begin{table}
    \begin{center}
        \begin{tabular}{lrcc}\hline
      & $\vec{w}_1$ & $\vec{w}_2$ & $\vec{w}_3 $  \\
        \hline
       $\vec{w}_1$& (b) & (a) &  (a) \\
     $\vec{w}_2$ &  (a) & (b) &  (a)\\
     $\vec{w}_3$ &  (a) &  (a) & (b)\\ \hline
        \end{tabular}
        \caption{Combinations} 
    \end{center}
\end{table}

 A simple symmetry in parameter space is the $(\zeta_1, \zeta_2) \to (-\zeta_1, -\zeta_2)$, as it does not change the decaying rates. In addition, from the table, we see that changing the region and the operation, interchanging the label of the former with the weight of later, gives a similar geometric phase behavior. In fact, there is an exact symmetry that can be discussed in terms of Weyl transformations $W_\alpha \in SU(3)$, where the label $\alpha$ denotes the roots of $\mathfrak{su}(3)$ 
(six possibilities, placed at $60^{\circ}$ along the borders of Fig. \ref{Regions}).
Using a root $\vec{\alpha}$, it is known that, 
\[
W_\alpha (\xi^q T_q) W_\alpha^\dagger = R^q(\xi)\, T_q \;,
\] 
where $R(\vec{\xi})$ is the reflection of $\vec{\xi} = (\xi_1, \xi_2)$ with respect to the line with perpendicular $\vec{\alpha}$,
\begin{equation}
R(\vec{\xi}) = \vec{\xi} - 2(\vec{\xi} \cdot \hat{a})\, \hat{a}   \makebox[.5in]{,}
\hat{a} = \vec{\alpha}/\sqrt{\vec{\alpha} \cdot \vec{\alpha}} \;.
\label{refle}
\end{equation}

If $H^{(i)}_{\rm S} = \vec{w}_i \cdot \vec{T}$, a Weyl transformation with $\vec{\alpha} = \alpha_{ij}$ leads to $ {H}^{(j)}_{\rm S}  = W_\alpha  H^{(i)}_{\rm S} W_\alpha^\dagger = \vec{w}_j \cdot \vec{T}$.  Then,
\begin{eqnarray}
\phi^{(i)}_g(t) & =& \arg \, {\rm Tr}  \left[ (\Psi'(0))^\dagger \,  U^{(i)}_{\rm S} \, \Psi'(t) \right]
 + \int_0^t ds \,\, {\rm Tr}  \left[ (\Psi'(s))^\dagger  H^{(i)}_{\rm S} \Psi'(s) \right]
 \nonumber \\
 &=& \arg \, {\rm Tr}  \left[ (\Psi'(0))^\dagger 
 W_\alpha^\dagger U^{(j)}_{\rm S}W_\alpha \Psi'(t) \right]
 + \int_0^t ds \, {\rm Tr}  \left[ (\Psi'(s))^\dagger   W_\alpha^\dagger U^{(j)}_{\rm S}W_\alpha \Psi'(s) \right] 
  \nonumber \\
 &=& \arg \, {\rm Tr}  \left[ (\Psi'_W(0))^\dagger  
 U^{(j)}_{\rm S} \Psi'_W(t) \right]
 + \int_0^t ds \,\, {\rm Tr}  \left[ (\Psi'_W(s))^\dagger U^{(j)}_{\rm S} \Psi'_W(s) \right] \;,
\end{eqnarray}
\begin{equation}
\Psi'_W(s) = W_\alpha \Psi'(t) W_\alpha^\dagger \;.
\end{equation}
This object, together with the transformed $\Psi'_1(t)$, $\Psi'_2(t)$, 
will be eigenvectors in a system where the $q$-components of $\tilde{\rho}$ get transformed by a reflection (cf. Eq. (\ref{m-not})). This in turn will be a solution to a master equation similar to Eq. (\ref{red-mas}), but with reflected components of the matrices $G_q$ or, equivalently,  
reflected coefficients $R_{q' q''}$ (cf. Eq. (\ref{g-prop})). Finally, this reflection is propagated to the couplings $(\zeta_1 , \zeta_2)$ via Eqs. (\ref{erres}), (\ref{gammas}) and (\ref{zetas}).

For example, performing a Weyl transformation with a root along the direction $\hat{a}=(\frac{1}{2}, \frac{\sqrt{3}}{2})$, the coupling $(0.3, 0.3)$ in the region labeled by 
$\vec{w}_3$ (dot in blue) is reflected to the coupling $-(0.11, 0.41)$ in the region labeled by $\vec{w}_1$ (dot in purple). The numerical evaluation of the geometric phase, for new coupling, with the system operated by $\vec{w}_3$, turned out to be that displayed in Fig. \ref{geo}, for the old coupling, with the system operated by $\vec{w}_1$ (blue line in plot (a)). In addition, the old asymptotic values for the effective state coefficients $(u_0, u_1, u_2)= (\sqrt{2/3}, -1/2,-1/\sqrt{12}) $ (Fig. \ref{asy-coef})  gave rise to new effective state coefficients $(\sqrt{2/3}, 0,1/\sqrt{3})$. Then, $u_0$ is unaltered, and the new $q$-sector $(0,1/\sqrt{3})$ is precisely $R(\vec{\xi})$ (cf. Eq. (\ref{refle})) with  
$\vec{\xi} = -(1/2, 1/\sqrt{12})$ (the old $q$-sector), as expected.

\section{Conclusions}
\label{conclusiones}

In this work, we studied the effect of the environment on a bipartite system formed by a pair of entangled qudits, focusing on the system's geometric phase and related physical quantities under dephasing. In many applications, due to the larger $d^2$-dimensional space of states, higher d-dimensional quantum states (qudits) could be more efficient than qubits.  Decoherence is the main obstacle to overcome since it is a process whereby the system loses its ability to exhibit quantum interference.  

Here, considering the weak coupling limit and disregarding dissipation, we analyzed the effect of noise on an initial state that is pure and close to a maximally entangled state (MES). For this aim, we derived a master equation relying on the two-qudit basis introduced in Ref. \cite{oxman1}, which contains the MES ($\sum_{i} |ii \rangle /\sqrt{d}$) as one of its elements. The remaining $d^2-1$ elements are in correspondence with the generators of the $\mathfrak{su}(d)$ Lie algebra, which can be separated into a $(d-1)$-dimensional Cartan sector and an off-diagonal sector. 
This description proved to be useful when proposing the fractional topological phase $2\pi/d $ generated in MES states operated by unitary evolutions \cite{oxman1}, and to understand the algebraic and topological aspects of two-qudit sectors with different concurrence \cite{Asp}. 

To analyze dephasing, the system and environment were coupled  through the diagonal degrees of each individual qudit (for qubits, this would correspond to the spin-component $\sigma_3$). Then, if the initial state is pure and restricted to the $d$-dimensional MES plus Cartan ``diagonal''   sector  ($|\psi(0)\rangle = \sum_{i} a_{ii} |ii \rangle $), we showed that the evolution of the reduced $d^2 \times d^2$ density matrix remains in this sector. That is, decoherence can be studied by means of a master equation for a $d\times d$ restricted density matrix $\tilde{\rho}$. The dynamics of the geometric phase is captured by an ``effective'' state, that is, the eigenstate  of the mixture $\tilde{\rho}$ which at $t=0$ coincides with the initial pure state. 
 
As an initial check, we re-obtained a known result. Namely, for a qubit pair in an initial MES state, the geometric phase is not affected by decoherence \cite{bipartite}.
Then, we turned our attention to qutrit pairs and the ``effective'' concurrence, as for locally operated isolated systems the geometric phase strongly depends on the (conserved) $I$-concurrence of the initial state. When the system is coupled to the environment, the effective concurrence $C(t)$ displays a nontrivial time dependence. Generally, it decays to zero at asymptotic times. However, an interesting behavior arises when the initial state approaches the MES state. In this case, $C(t)$ is more protected against the effects of the environment. For given moderate couplings, when the initial state gets closer to the MES state, we clearly see that the effective concurrence is stabilized around its initial value for more periods. This is due to the emergence of a softer destabilization regime, as compared with the asymptotic decay to zero. A ``kink'' localized in time arises, representing an initial decay to a nonzero value, then an increase up to $C(0)$, followed by a decay to zero at later times. The peak in $C(t)$ occurs when a pair of eigenvalues of the density matrix get closer, with associated eigenvectors  $\approx (|22\rangle + | 33\rangle ) /\sqrt{2}$ and
$| 11\rangle $. They form a quasi-degenerate subspace,  allowing for the
transient formation of a state close to the MES, 
$ (\left| 11\rangle + | 22\rangle\ + | 33\rangle\right)/\sqrt{3}$. Furthermore, as the initial concurrence is increased, the kink moves to larger times and, when the MES is reached, only the decay to the nonzero value is left. This value happens to be $1$; accordingly, we verified that the environment drives the initial two-qutrit MES into an effective two-qubit MES state.  

Finally, we studied the geometric phase for an initial MES, operating the system with three different evolutions. They correspond to the three possible pairs of coefficients (weights) used to combine the Cartan generators. If the system were isolated, they would generate the fractional topological phase.  In addition, the system-environment interaction is described by a pair of couplings, which can be divided into three regions. The effect of decoherence, depends on which combination of operating weight and coupling region is realized, with the different physical quantities (operations, GPs, couplings, effective state coefficients) displaying an exact Weyl symmetry. Among the nine possibilities, the zero-coupling pattern is destroyed in three of them. In the other six cases, the geometric phase gradually moves from an initial regime with jumps of $2\pi/3$ to a later regime with jumps of $\pi$, in accordance with the MES qutrit-to-qubit transition driven by the environment. This, together with the enhanced stability properties around the maximally entangled state, delineates a strategy to minimize the effects of the environment on fractional topological phases. Further research on this direction will be presented elsewhere.

 


\section*{Acknowledgements}
L. E. O.  and A. Z. K. would like to acknowledge the
Conselho Nacional de Desenvolvimento Tecnol\'ogico (CNPq), 
Coordena\c c\~{a}o de Aperfei\c coamento de 
Pessoal de N\'\i vel Superior (CAPES), Funda\c c\~{a}o de Amparo \`{a} 
Pesquisa do Estado do Rio de Janeiro (FAPERJ-BR), and Instituto Nacional 
de Ci\^encia e Tecnologia de Informa\c c\~ao Qu\^antica (INCT-CNPq) for financial support. 
F. C. L. and P. I. V. are supported by ANPCyT, CONICET, and UBA.


\begin{thebibliography}{99}
\bibitem{footnote} A tuple is defined as positive if the last nonvanishing component is positive.
\end{thebibliography}
\end{document}